# Correlation factor for diffusion in cubic crystals with solute-vacancy interactions of arbitrary range

*J.L. Bocquet*
*CMLA, LRC-Méso, ENS CACHAN, 61 Avenue du Président Wilson, 94235 Cachan, France*
*jean-louis.bocquet@cmla.ens-cachan.fr*

## Introduction

Random walk theory is at the root of an atomic description of matter transport in the solid state. Any particle following a path in which the direction of a step is independent of the direction of the preceding one is a random walker. In a crystal lattice (with coordination number z) the sites of which are occupied by a single atomic species, a vacancy performing first neighbour jumps follows a random walk because the probability of performing a jump in any direction is always 1/z. Its average quadratic displacement after 'n' jumps of length 's' is given by the random walk result [1] $\langle R^2 \rangle = ns^2$.

Conversely a substitutional atom occupying a lattice site is not a random walker: most of the time it can move only if it exchanges its position with a first neighbour vacancy. After a first exchange jump along some direction, the vacancy sits in the proper location to make the same atom perform immediately a reverse jump. From the point of view of the atom, this backward jump has a probability of occurrence different from 1/z and larger than that attached to a jump in any other direction. This memory effect, due to the geometry of the jump mechanism, induces a correlation between the directions of two successive jumps of the same atom: this is the reason why the pure random walk theory had to be corrected.[2] The average quadratic displacement of the atom after 'n' jumps is written as $\langle R^2 \rangle = f\ ns^2$. The correlation factor $f$ is a correction for the non-randomness of the displacements and is always smaller than unity for a vacancy mechanism.

An abundant literature has been devoted to the evaluation of correlation effects since the fifties of the last century: a lot of approximate or exact results were produced, according to the methods at work and the cases which were examined for various diffusion mechanisms, namely, vacancy, interstitial and intersticialcy ones (see for instance a review article [3] giving an excellent survey of the state of the art at that time).

The most simple case ever studied is the diffusion of a tagged atom called 'the tracer' moved by a single vacancy in an infinite crystal lattice. The chemical nature of the tracer can be identical ('self diffusion') or different ('solute diffusion') from that of all other atoms. In both cases the calculation of the correlation effect boils down to evaluate the return probabilities $p_{\omega_1,\omega_2}$ of performing a tracer jump $\overrightarrow{\omega_2}$ following a previous jump $\overrightarrow{\omega_1}$, with the help of the *same* vacancy. The average quadratic displacement of the tracer due to its encounter with a given vacancy is a function of the number of jumps and of the probabilities $p_{\omega_1,\omega_2}$. Due to the translational symmetry of the medium, the atomic configuration after a tracer jump is identical to that before the jump, apart from a rotation: the probabilities $p_{\omega_1,\omega_2}$ do not depend on $\overrightarrow{\omega_1}$ and $\overrightarrow{\omega_2}$ separately, but only on $\overrightarrow{\omega_2} - \overrightarrow{\omega_1}$. In such an ideal situation, a simple recurrence equation for the displacement of the tracer atom can be written and solved in order to extract the mean quadratic displacement of the tracer due to its encounter with a single vacancy. One finds easily:

$$\langle R^2 \rangle = \frac{s^2}{1-P}\frac{1+Q}{1-Q} \qquad P = \sum_{\omega_2 = \{\omega_{1V}\}} p_{\omega_1,\omega_2} \qquad Q = \sum_{\omega_2 = \{\omega_{1V}\}} p_{\omega_1,\omega_2} \cos(\overrightarrow{\omega_1}, \overrightarrow{\omega_2}), \quad (1)$$

where $P$ is the total probability of performing the next tracer jump and $Q$ the average cosine between two successive tracer jumps; the summation runs over the set of first neighbour jump vectors $\{\omega_{1V}\}$.

The probability $P$ is smaller than unity for 3D lattices. The total number of tracer jumps performed with the help of the same vacancy is $\langle n \rangle = 1 + P + P^2 + ... = 1/(1-P)$ and the random contribution to the quadratic displacement of the tracer is given by $\langle R^2 \rangle_{rand} = \langle n \rangle s^2 = s^2/(1-P)$. The departure from the random case is the correlation factor given by :

$$f = \langle R^2 \rangle / \langle R^2 \rangle_{rand} = (1+Q)/(1-Q). \quad (2)$$

At the start, only approximate values of the return probabilities were available, taking into account only a limited number of vacancy jumps.[2] A further extension took into account all the trajectories made of an arbitrarily large number of defect jumps, provided they were entirely contained in a small spherical volume V around the tracer (matrix methods [4]): the implicit assumption was that a vacancy stepping out of V through some exit site was considered to be lost for the correlation problem; as a consequence any further return into V was considered as independent from the location of the exit site. The most interesting feature of the matrix methods was the production of analytical expressions of the correlation factor as a function of the vacancy jump frequencies in volume V. For spare of tractability however, the volume V was always restricted to the first and, in some cases, second neighbour shells. Further improvements (extended matrix methods [5]) took an approximate account of defect trajectories outside V.

On the other hand a small number of exact values for correlation factors were known as by-products of the random walk theory (see an excellent and thorough review [6]): namely, for diffusion by the vacancy mechanism in a 2D square lattice $f = 1/(\pi - 1)$.[7] A clever electric analogy of the atomic transport with the vacancy mechanism found also $f = 1/3$ in the 2D honeycomb lattice and $f = 1/2$ in the 3D diamond lattice.[8]

Integral methods taking explicitly into account the whole lattice appeared only later; although they rest on different treatments of the transport equation, namely Green functions,[9] double Laplace and Fourier transforms,[10] random walk generating functions,[11-12] they yielded as expected the same final result as a function of elliptic integrals. If some of them are known analytically, the majority has of course to be evaluated numerically. Former exact results were recovered together with the discovery of a new explicit value of the correlation factor in a 2D triangular lattice $f = (\pi + 6\sqrt{3})/(11\pi - 6\sqrt{3})$.[9] The interest of integral methods is to yield exact values of correlation factors and to express the numerical coefficients entering the approximate expressions established earlier as combinations of lattice integrals. The exact values of correlation factors obtained in this way are systematically lower than all those obtained previously: indeed, dropping the contribution of defect trajectories which leave the volume V produces an under-estimation of the correlation bias, hence an over-estimation of the correlation factor. But in many cases, they confirmed the quality of the results previously obtained by the extended matrix methods.

Looking back over the path followed during the half past century, a considerable evolution shows up today. Indeed, the way of tackling the problem of matter transport in solids is undergoing a complete reversal.

* Formerly, macroscopic diffusion experiments together with direct measurements of vacancy concentrations were the only way to gain some information about the microscopic description of the jump mechanism and jump frequencies at atomic scales: they were smartly combined to deduce properties of the metallic systems which were out of reach of direct measurements.[13-18] As a result, values of vacancy jump frequencies in the close neighbourhood of solute atoms were proposed for a number of FCC [19] and BCC [20] metallic binary systems. However the experiments quoted above did not yield the jump frequencies themselves, but only a few ratios of them. Indeed the measured quantities are not fully independent, a fact which can be traced back in their analytical expressions which use the same combinations of frequencies. As a result, even the simplest models had frequently to be downgraded with further simplifying assumptions to make their best from the small number of independent measurements.

* Today, the improvement of ab-initio methods and the availability of computational codes have definitively converted sophisticated calculations into (nearly) routine tasks. The atomic parameters, which were formerly the ultimate quest of macroscopic measurements, are now calculated with a rather high level of confidence and have taken the place of the input data to be plugged into the modelling, which is in charge of evaluating macroscopic fluxes during phase transformations. This trend reversal enlightens the fact that the today available analytical models have to be extended in order to incorporate a thorough atomic description of transport. For instance the modelling of the past was never formulated for a solute-vacancy interaction range extending beyond the second neighbour shell, for sake of analytical tractability; recent ab-initio results did show however that solute-vacancy interactions extend often well beyond,[21] or exhibit a pattern of frequencies which cannot be inserted into the simplifying assumptions of previous modelling.[22]

The present contribution addresses the problem of tracer solute diffusion via a vacancy mechanism in FCC and BCC lattices and calculates the return probabilities with integral methods for an arbitrary range of vacancy-solute interactions. The paper is organized as follows:

* the first part is devoted to recall the main features of the method at work, which was illustrated in previous publications;[10,23]

* the second part shows how to reduce the size of the problem with the help of symmetries;

* the third part examines the case of the so-called five-frequency model in a FCC lattice as a benchmark test;

* the fourth part proposes a formal expression of the solution for an arbitrarily large interaction range. Apart from numerical coefficients given by lattice integrals, which can be calculated once for all, the final step is the numerical solution of a system of linear equations. Model cases are examined, encompassing oscillating of monotonous interactions over longer distances than those studied before;

* the fifth part applies the general method to systems recently studied;

* appendices deal with details of analytical calculations and/or numerical results. Appendix A gives some hints on the general formalism. Appendix B reformulates an exact result previously obtained for the so-called five frequency model. Appendix C gives a brief description of the computational Fortran code which has been used throughout this work. Appendix D details relationships between lattice integrals for FCC and BCC lattices.

## 1. The method at work

The method consists in writing down a transport equation for the function $L(r,t)$ which stands for the probability of finding the vacancy on site r at time t. In the bulk, i.e. far from the tracer atom, the time derivative is simply given by a balance between ingoing and outgoing contributions to the vacancy probability on site r:

$$\frac{dL(r,t)}{dt} = -12W_O L(r,t) + W_O \sum_{\{\omega_{1V}\}} \left[ L(r+\omega_{1V},t) \right], \qquad (3)$$

where $W_O$ stands for the vacancy jump frequency in the bulk material and the summation runs over the first neighbour jump vectors $\{\omega_{1V}\}$.

This general expression must then be corrected in order to take into account *specific sites*, i.e. sites with properties departing from those of a bulk site:

* the origin denoted hereafter $R_0$ is a special site which plays the role of a sink: any trajectory of the vacancy passing through the origin lets the tracer perform a second jump and makes an end of its contribution to the correlation effect. Therefore the outgoing jump frequency from the origin is set to zero and the first neighbour sites of the origin receive no contribution from the origin;

* for all lattices sites lying inside the range of the solute-vacancy interaction, some (or all) of the outgoing or ingoing frequencies depart from $W_O$. These sites, named above *specific sites*, are denoted by $R_i$ (i=1,N). For each specific site (origin included), a $\delta()$ operator is introduced to add a corrective term which modifies the general balance written above;

* the vacancy starts from site $-\overrightarrow{\omega_1}$ which is a first neighbour site of the origin occupied by the tracer atom, since this site is the location of the vacancy after its first exchange with the tracer. The initial condition of the transport equation reads then $L(r,0) = \delta(r+\omega_1) = ci(r)$.

The complete transport equation reads now:

$$\begin{aligned} \frac{dL(r,t)}{dt} &= -12W_O L(r,t) + W_O \sum_{\{\omega_{1V}\}} \left[ L(r+\omega_{1V},t) \right] \\ &+ \sum_{\{R_j\}} \delta(r-R_j) \sum_{\{\omega_{1V}\}} \delta(r+\omega_{1V}-\{R_k\}) \left[ L(r,t)(W_O - W_{j\to k}) + L(r+\omega_{1V},t)(W_{k\to j} - W_O) \right] \end{aligned} \qquad (4)$$

The second line includes all the corrections for specific sites and uses two summations in series:

* the first summation scans all the specific sites;

* the second summation runs over the first neighbours of the specific site $R_j$ : if the vector $R_j + \omega_{1V}$ points at whichever specific sites $R_k$, the outgoing (ingoing) frequency from (to) site $R_j$ is different from $W_O$ and must be replaced by $W_{j\to k}$ ($W_{k\to j}$)

which stands for the jump frequency from the specific site $R_j$ ( $R_k$ ) to the specific site $R_k$ ( $R_j$ ). This frequency is set to zero if $R_j$ and $R_k$ are not first neighbours or if the starting site of the jump is the origin. Due to the constraint imposed by the two $\delta()$ operators in series, the internal summation runs only on those sites for which at least one of the outgoing or ingoing frequency differs from $W_O$.

### 1.1 Laplace and Fourier transforms of the transport equation

The sought after return probabilities can then be written as time integrals:

$$p_{\omega_1,\omega_2} = W_{\omega_{1V}\to 0}\int_0^\infty L(\omega_2,t)dt\,, \tag{5}$$

where $W_{\omega_{1V}\to 0}$ is the vacancy jump frequency from a first neighbour $\overrightarrow{\omega_{1V}}$ onto the origin. This probability is nothing but the Laplace transform $LL(r,p)$ of $L(r,t)$ calculated at $r=\omega_2$ for $p=0$, since :

$$LL(r,p)=\int_0^\infty L(r,t)e^{-pt}dt \text{ and } p_{\omega_1,\omega_2}=W_{\omega_{1V}\to 0}\,LL(\omega_2,p)\big|_{p=0}\,. \tag{6}$$

We define the Fourier transform and its reverse by

$$FL(k,t)=\sum_{\{r\}} e^{-ikr}L(r,t) \Leftrightarrow L\,(r,t)=\frac{1}{V_{ZB}}\int_{V_{ZB}} e^{+ikr}FL(k,t)d_3k\,, \tag{7}$$

where the exponent $kr$ stands for a scalar product, the summation runs over all lattice sites and $V_{ZB}$ is the volume of the first Brillouin zone. The Fourier transform of the initial condition is $ci(k)=e^{ik\omega_1}$.

In what follows, we will also have to make use of the identity :

$$\sum_{\{r\}} e^{-ikr}L(r+\omega,t)=\sum_{\{r\}} e^{-ik(r+\omega)}L(r+\omega,t)e^{ik\omega}=FL(k,t)e^{ik\omega}\,.$$

The double Fourier and Laplace transform is defined by two equivalent expressions:

$$FLL(k,p)=\sum_{\{r\}} e^{-ikr}LL(r,p)=FL(k,t)\big|_{t=0}+\int_0^\infty FL(k,t)e^{-pt}dt\,, \tag{8}$$

and the reverse Fourier transform applied to the doubly transformed function reads :

$$LL(r,p)=\frac{1}{V_{ZB}}\int_{V_{ZB}} e^{+ikr}FLL(k,p)d_3k\,. \tag{9}$$

After performing the Fourier and Laplace transforms, Eq. (4) becomes :

$$pFLL(k,p)-ci(k)=-FLL(k,p)Deno+\sum_{\{R_j\}}e^{-ikR_j}$$
$$\left[\sum_{\{\omega_{1V}\}}\delta(R_j+\omega_{1V}-\{R_k\})LL_j(W_O-W_{j\to k})+\sum_{\{\omega_{1V}\}}\delta(R_j+\omega_{1V}-\{R_k\})LL_k(W_{k\to j}-W_O)\right], \tag{10}$$

where $Deno=W_O(z-\sum_{\{\omega_{1V}\}} e^{ik\omega_{1V}})$ and $LL_j=LL(R_j,p)$. (11)

It is worth noticing that none of these quantities $LL_j$ will raise any problem if $p\to 0$, but that attached to the origin, namely $LL_0$. Indeed, the probability which accumulates on the origin is a monotonously increasing function of time with an horizontal asymptote. Its Laplace transform behaves like $1/p$ and will require a special handling as explained later.

### 1.2 Formal expression of the return probabilities

The doubly transformed function $FLL(k,p)$ is easily extracted from Eq. (10) and its reverse Fourier transform yields the probabilities on any desired site $R_l \quad l=0,N$ through Eq. (9) while setting $p=0$:

$$LL_l = \frac{1}{V_{ZB}} \int_{V_{ZB}} \frac{ci(k) e^{+ikR_l}}{Deno} d_3k + \sum_{\{R_j\}} \frac{1}{V_{ZB}} \int_{V_{ZB}} \frac{e^{+ik(R_l - R_j)}}{Deno} d_3k \left[ \sum_{\{\omega_{1V}\}} \delta(R_j + \omega_{1V} - \{R_k\}) \left( LL_j (W_O - W_{j\to k}) + LL_k (W_{k\to j} - W_O) \right) \right]. \quad (12)$$

Using the shortened notations

$$I_{R_j} = \frac{1}{V_{ZB}} \int_{V_{ZB}} \frac{e^{+ikR_j}}{Deno} d_3k \qquad I_{ciR_j} = \frac{1}{V_{ZB}} \int_{V_{ZB}} \frac{ci(k) e^{+ikR_j}}{Deno} d_3k \, , \quad (13)$$

we obtain a set of N linear equations which reads

$$LL_l - \sum_{\{R_j\}} \left[ \sum_{\{\omega_{1V}\}} \delta(R_j + \omega_{1V} - \{R_k\}) I_{R_l - R_j} \left( LL_j (W_O - W_{j\to k}) + LL_k (W_{k\to j} - W_O) \right) \right] = I_{ciR_l} \, , \quad (14)$$

where index $l$ runs over the N specific sites.

The attention of the reader is drawn on the fact that the above formulation did not gather together all the contributions entering the coefficient of the quantity $LL_l$, and, as is, cannot be used straightforwardly to solve the linear system. In addition a further handling of the double summation has to be performed. We show in Appendix A that the presence of the unknown $LL_O$, which diverges when $p \to 0$, does not raise any difficulty when solving the linear system.

## 2. Reduction of system size thanks to symmetries

The number N of unknowns is a very rapidly increasing function of the solute-vacancy interaction range: for a 1$^{st}$, 2$^{nd}$, 3$^{rd}$, 4$^{th}$, 5$^{th}$-neighbour interaction range, the number N of sites to be taken into consideration a priori is 54, 78, 134, 176, 200 in a FCC lattice, and 34, 50, 88, 136, 168 in a BCC one, respectively. Symmetries must be used to cut down the number of unknowns:

* we define the crystal lattice sites by: $r_{ix+jy+kz} = i\,\Omega_x + j\,\Omega_y + k\,\Omega_z$ with a set of basis vectors $\Omega_x = a/2[1,0,0] \qquad \Omega_y = a/2[0,1,0] \qquad \Omega_z = a/2[0,0,1]$ where « a » stands for the lattice parameter and i,j,k are integers; i+j+k must be even for FCC; i,j,k must be all odd or all even for BCC. We will use hereafter the simplifying notation

$$c_{lx+my+nz} = \cos(lk_x + mk_y + nk_z) \qquad s_{lx+my+nz} = \sin(lk_x + mk_y + nk_z) \quad (15)$$

where $k_x$, $k_y$, $k_z$ are the vector components in reciprocal space and where indices 'l', 'm' or 'n' are omitted whenever equal to unity. The expressions of $Deno$ for a FCC and BCC lattice become respectively

$$Deno = W_O(12 - 4c_x c_y - 4c_y c_z - 4c_z c_x) = W_O D_0 \quad (16)$$

$$Deno = W_O(8 - 8c_x c_y c_z) = W_O D_0 ; \tag{17}$$

* instead of calculating the correlation factor along an arbitrary axis, we evaluate it along a principal axis of the diffusion tensor.[25] For cubic lattices, a convenient principal axis is <100>, taken afterwards as the x-axis. As a consequence only the x-components of the jump vectors are to be considered. With our choice of unit vectors, the average cosine $Q$ introduced above in Eq. (1) is replaced by the average product $< x_1 x_2 >$ of the x-components of two successive tracer jumps: $Q$ is evaluated by summing with a negative (positive) sign the contributions brought by the occupancy probabilities of sites located on the positive (negative) side of the x-axis;

* thanks to the linearity of the transport equation, the initial condition on site $-\overrightarrow{\omega_1}$ in the general formulation can be replaced by a positive unit source equally split over the four first neighbour sites with positive abscissas, together with a negative unit source split over the first neighbour sites with negative abscissas. This new initial condition transfers the four-fold symmetry of the cubic lattice to the diffusion problem itself.

For a FCC lattice, the initial condition and its Fourier transform are then given by

$$\left.\begin{aligned} L(r,t=0) = &\frac{1}{4}\left[\delta(r-\omega_{110}) + \delta(r-\omega_{101}) + \delta(r-\omega_{1\bar{1}0}) + \delta(r-\omega_{10\bar{1}})\right] \\ &-\frac{1}{4}\left[\delta(r-\omega_{\bar{1}10}) + \delta(r-\omega_{\bar{1}01}) + \delta(r-\omega_{\bar{1}\bar{1}0}) + \delta(r-\omega_{\bar{1}0\bar{1}})\right] \end{aligned}\right\} \rightarrow ci(k) = -is_x(c_y + c_z) \tag{18}$$

$ci(k) = -iCI(k)$

For a BCC lattice: i,j,k must be all odd or all even. The initial condition, its Fourier transform and the expression of $Deno$ are given by

$$\left.\begin{aligned} L(r,t=0) = &\frac{1}{4}\left[\delta(r-\omega_{111}) + \delta(r-\omega_{1\bar{1}1}) + \delta(r-\omega_{11\bar{1}}) + \delta(r-\omega_{1\bar{1}\bar{1}})\right] \\ &-\frac{1}{4}\left[\delta(r-\omega_{\bar{1}11}) + \delta(r-\omega_{\bar{1}\bar{1}1}) + \delta(r-\omega_{\bar{1}1\bar{1}}) + \delta(r-\omega_{\bar{1}\bar{1}\bar{1}})\right] \end{aligned}\right\} \rightarrow ci(k) = -is_x(2c_y c_z) \;. \tag{19}$$

$ci(k) = -iCI(k)$

As a consequence, the plane x=0 becomes a mirror symmetry plane, all sites of which (origin included) have a zero occupancy probability. All the unknowns attached to specific sites located on plane x=0 disappear from the equations, $LL_O$ included. In the same way the occupancy probabilities on mirror sites have opposite signs $LL_i^+ = -LL_i^-$.

* due to the four-fold symmetry of the x-axis for the diffusion problem, the specific sites can thus grouped into $M_{sub}$ subsets $S_i^+$ or $S_i^-$ of equal occupancy probability $LL_i^+$ or $LL_i^-$ and numbered in ascending order, according to their x-coordinate and their distance to the x-axis ; the superscript is '+' or '-' depending on whether they are on the positive or negative x side ; no superscript is used for the subsets in plane x=0. The sets of vectors defining the defective sites are still denoted $\{R_i^+\}$, $\{R_i\}$ and $\{R_i^-\}$, but now with subscript $i = 0, M_{sub}$. The defective sites in each subset will also be represented by the *representative site* of the subset; its coordinates $(i_1, i_2, i_3)$ comply with the condition $i_1 > 0,\ i_2 \geq 0,\ i_3 \geq 0$ and $i_2 \leq i_3$ for $\{R_i^+\}$, $i_1 = 0$ , $i_2 \geq 0$ , $i_3 \geq 0$ and $i_2 \leq i_3$ for $\{R_i\}$, $i_1 < 0$ , $i_2 \geq 0$ , $i_3 \geq 0$ and $i_2 \leq i_3$ for

$\{R_i^-\}$. Knowledge of the triplet $(i_1, i_2, i_3)$ is sufficient to generate the other members of the subset, through sign changes and interchange of $i_2$ and $i_3$. The only unknowns to be considered will then be $LL_i^+$, which decreases the number of equations to be solved from N down to $M_{sub}$ and restricts the only subsets to be considered to those with a positive abscissa;

* since the first subsets located in plane x=0 have a zero occupancy probability, they have no part in the final result; the summation on the relevant subsets will ten run from $m_{sub}$ to $M_{sub}$, where $m_{sub}$ is the number attached to the first subset with non-zero x-component. For the vacancy mechanism presently studied, the average product $< x_1 x_2 >$ will bring into play only the occupancy probability on subset with number $m_{sub}$ and will be expressed by

$$Q = -< x_1 x_2 >= -n_{m_{sub}} W_{\omega_{1V} \to 0} LL^+_{m_{sub}} \,, \qquad (20)$$

where $n_{m_{sub}}$ is the number of sites in subset $m_{sub}$ on the positive x-side.

As a consequence of the above remarks, the correction terms for sites belonging to the same subset bring an equal contribution to the Fourier transform. They make simple functions $f_l^+$, $f_l^-$ come out in the final formulation, together with their difference $f_l = f_l^+ - f_l^-$ :

$$\left.\begin{aligned} f_l^+ &= \sum_{\{R_l^+\}} e^{-ikR_l^+} = e^{-ik_x l_1} \sum_{\{l_2, l_3\}_l} e^{-ik_y l_2 - ik_z l_3} \\ f_l^- &= \sum_{\{R_l^+\}} e^{-ikR_l^-} = e^{+ik_x l_1} \sum_{\{l_2, l_3\}_l} e^{-ik_y l_2 - ik_z l_3} \end{aligned}\right\} \to f_l = -2is_{l_1 x} \sum_{\{l_2, l_3\}_l} e^{-ik_y l_2 - ik_z l_3} = -iF_l \,, \qquad (21)$$

where the summations run only on the components $\{l_2, l_3\}_l$ of the vectors belonging to $\{R_l^+\}$.

Their analytic expression depends on the value of $l_2, l_3$ and reads formally:

$$\begin{aligned} \sum_{\{l_2, l_3\}_l} e^{-ik_y l_2 - ik_z l_3} &= \delta_{l_2 - l_3} \left[ \delta_{l_2} \delta_{l_3} + (1 - \delta_{l_2} \delta_{l_3}) 4 c_{l_2 y} c_{l_2 z} \right] \\ &+ (1 - \delta_{l_2 - l_3}) \left[ \delta_{l_2} (2c_{l_3 y} + 2c_{l_3 z}) + \delta_{l_3} (2c_{l_2 y} + 2c_{l_2 z}) + (1 - \delta_{l_2} - \delta_{l_3}) \, (4c_{l_2 y + l_3 z} + 4c_{l_3 y + l_2 z}) \right] \end{aligned} \qquad .(22)$$

Grouping altogether the correction terms which bring contributions of equal magnitude and taking advantage of the mirror symmetry plane, the transport equation becomes

$$(p + Deno) FLL(k, p) = ci(k) + \sum_{j = m_{sub}}^{M_{sub}} f_j \sum_{kVj} nlien_{j \to k} \left[ LL_j^+ (W_O - W_{j \to k}) + LL_k^+ (W_{k \to j} - W_O) \right],$$

where the first summation runs now on the number of subsets and the second summation with index $kVj$ running only on those subsets ‘k’ which can be reached from subset ‘j’ in one jump. This neighbourhood relationship between subsets is stored in a variable $nlien_{j \to k}$ which is the number of sites of subset ‘k’, which can be reached in one jump from a given site of subset ‘j’. Hence the expression of the doubly transformed function where $p$ has been set to zero:

$$FLL(k,p)\big|_{p=0} = \frac{ci(k)}{W_O D_0} + \sum_{j=m_{sub}}^{M_{sub}} \frac{f_j}{D_0} \sum_{kVj} nlien_{j\to k}\left[ -LL_j^+ \, W'_{j\to k} + LL_k^+ \, W'_{k\to j} \right], \tag{23}$$

and where reduced frequencies $W'_{j\to k} = (W_{j\to k} - W_O)/W_O$ were used.
Grouping altogether the terms related to each unknown yields the formal result

$$FLL(k,p)\big|_{p=0} = \frac{ci(k)}{W_O D_0} + \sum_{j=m_{sub}}^{M_{sub}} LL_j^+ \frac{g_j(k)}{D_0}, \tag{24}$$

where the $g_j(k)$ are linear combinations of the functions $f_i(k)$.

At last we also take advantage of this more compact formulation to evaluate the desired return probabilities. Instead of using the inverse Fourier transform on only one site of subset $\{R_j^+\}$ to get

$$LL_j^+ = \frac{1}{V_{ZB}} \int_{V_{ZB}} e^{+ikR_j^+} \, FLL\,(k,p)\big|_{p=0} \, d_3k\,, \tag{25}$$

we perform the average on all the sites belonging to $\{R_j^+\}$ and $\{R_j^-\}$, taking due account of the sign change when switching from $R_j^+$ to $R_j^-$. We obtain analytical expressions where the space variables 'y' and 'z' play the same role and which bring into play the same functions $f_j^+$ and $f_j^-$ as those defined above

$$\begin{aligned} LL_j^+ &= \frac{1}{V_{ZB}} \int_{V_{ZB}} \frac{\sum_{\{R_j^+\}} e^{+ikR_j^+} - \sum_{\{R_j^-\}} e^{+ikR_j^-}}{(n_j^+ + n_j^-)} FLL\,(k,p)\big|_{p=0} \, d_3k \\ &= -\frac{1}{n_j V_{ZB}} \int_{V_{ZB}} f_j(k) FLL\,(k,p)\big|_{p=0} \, d_3k \end{aligned}, \tag{26}$$

where $n_j^+$ , $n_j^-$ stand for the number of sites in subsets $R_j^+$, $R_j^-$ and $n_j = n_j^+ + n_j^-$.
Hence the final system of $M_{sub} - m_{sub} + 1$ equations to be solved

$$\sum_{j=m_{sub}}^{M_{sub}} \left\{ \delta_{ij} - M_{ij} \right\} LL_j^+ = M_i \qquad i = (m_{sub}, M_{sub}), \tag{27}$$

with $M_{ij} = -\frac{1}{n_i V_{ZB}} \int_{V_{ZB}} \frac{f_i(k) g_j(k)}{D_0} d_3k \qquad M_i = -\frac{1}{n_i V_{ZB}} \int_{V_{ZB}} f_i(k) \frac{ci(k)}{W_O D_0} \, d_3k\,.$

## 3. Illustration on the case of the five frequency model for the FCC lattice

The range of the vacancy-solute interaction is limited to the 1$^{rst}$ neighbour shell. A site of the 1$^{rst}$ neighbour shell is first neighbour of sites in the 4$^{th}$ neighbour shell at most. The subsets must thus be defined for sites belonging to the first 4 neighbour shells.

### 3.1 Defining the subsets

Sites belonging to the mirror plane are considered first and grouped into the subsets $R_0, \{R_1\}, \{R_2\}, \{R_3\}$. Further subsets are numbered in increasing order, according to their abscissa, while scanning the shells of neighbours up to the 4$^{th}$:

* the intersections of the 1st neighbour shell with planes x=-1, x=0 and x=1 give rise to the new subsets $\left\{R_4^-\right\},\left\{R_4^+\right\}$ ; hence $m_{sub}=4$ $\quad n_{m_{sub}}=n_4^+$;

* the intersections of the 2nd neighbour shell with planes x=-2, x=0 and x=+2 give rise to the new subsets $\left\{R_5^-\right\},\left\{R_5^+\right\}$;

* the intersections of the 3rd neighbour shell with planes x=-2, x=-1, x=0, x=+1, x=+2 give rise to the new subsets $\left\{R_7^-\right\},\left\{R_6^-\right\},\left\{R_6^+\right\},\left\{R_7^+\right\}$ ;

* the intersections of the 4th neighbour shell with planes x=-2, x=0, x=+2 gives rise to the new subsets $\left\{R_8^-\right\},\left\{R_8^+\right\}$. Hence $M_{sub}=8$.

The mirror symmetry together with the four-fold axis induces the following relations among the unknowns: $LL_i=0$ $\quad (i=0,3)$ $\quad LL_i^-=-LL_i^+$ $\quad (i=4,8)$.

The five unknowns to be determined are $LL_4^+$ $LL_5^+$ $LL_6^+$ $LL_7^+$ $LL_8^+$ and the value of the average cosine is expressed by $Q=-4W_2LL_4^+$.

All the subsets and their definitions are gathered together in Table 1 below.

| Subset $S_i$ | Sites $\left\{R_i\right\}=\left\{R_i^+\right\}\cup\left\{R_i^-\right\}$ | $f_i=f_i^+-f_i^-=-iF_i$ | $n_i=n_i^++n_i^-$ |
|---|---|---|---|
| 0 | $R_0=0$ | 0 | 1 |
| 1 | $\left\{R_1\right\}=\left\{r_{011},r_{01\bar{1}},r_{0\bar{1}1},r_{0\bar{1}\bar{1}}\right\}$ | 0 | 4 |
| 2 | $\left\{R_2\right\}=\left\{r_{020},r_{002},r_{0\bar{2}0},r_{00\bar{2}}\right\}$ | 0 | 4 |
| 3 | $\left\{R_3\right\}=\left\{r_{022},r_{0\bar{2}2},r_{02\bar{2}},r_{0\bar{2}\bar{2}}\right\}$ | 0 | 4 |
| 4 | $\left\{R_4^+\right\}=\left\{r_{110},r_{101},r_{1\bar{1}0},r_{10\bar{1}}\right\}$<br>$\left\{R_4^-\right\}=\left\{r_{\bar{1}10},r_{\bar{1}01},r_{\bar{1}\bar{1}0},r_{\bar{1}0\bar{1}}\right\}$ | $-2is_x(2c_y+2c_z)$ | 8 |
| 5 | $\left\{R_5^+\right\}=\left\{r_{200}\right\}$<br>$\left\{R_5^-\right\}=\left\{r_{\bar{2}00}\right\}$ | $-2is_{2x}$ | 2 |
| 6 | $\left\{R_6^+\right\}=\left\{r_{121},r_{12\bar{1}},r_{1\bar{2}1},r_{1\bar{2}\bar{1}},r_{112},r_{11\bar{2}},r_{1\bar{1}2},r_{1\bar{1}\bar{2}}\right\}$<br>$\left\{R_6^-\right\}=\left\{r_{\bar{1}21},r_{\bar{1}2\bar{1}},r_{\bar{1}\bar{2}1},r_{\bar{1}\bar{2}\bar{1}},r_{\bar{1}12},r_{\bar{1}1\bar{2}},r_{\bar{1}\bar{1}2},r_{\bar{1}\bar{1}\bar{2}}\right\}$ | $-2is_x(4c_{2y}c_z+4c_yc_{2z})$ | 16 |
| 7 | $\left\{R_7^+\right\}=\left\{r_{211},r_{21\bar{1}},r_{2\bar{1}1},r_{2\bar{1}\bar{1}}\right\}$<br>$\left\{R_7^-\right\}=\left\{r_{\bar{2}11},r_{\bar{2}1\bar{1}},r_{\bar{2}\bar{1}1},r_{\bar{2}\bar{1}\bar{1}}\right\}$ | $-2is_{2x}(4c_yc_z)$ | 8 |
| 8 | $\left\{R_8^+\right\}=\left\{r_{220},r_{202},r_{2\bar{2}0},r_{20\bar{2}}\right\}$<br>$\left\{R_8^-\right\}=\left\{r_{\bar{2}20},r_{\bar{2}02},r_{\bar{2}\bar{2}0},r_{\bar{2}0\bar{2}}\right\}$ | $-2is_{2x}(2c_{2y}+2c_{2z})$ | 8 |

*Table 1. Definition of the subsets entering the five frequency model.*

### 3.2 Defining and naming the vacancy jump frequencies

The link between the present notation, which uses jump frequencies between subsets, and the standard formulation which uses jump frequencies between neighbour shells, must be established. It is detailed below in Table 2. The first four lines pertain to the four subsets contained in plane x=0; the next five ones pertain to subsets on the positive side of the x-axis.

| $W_{i\to j}$ ($i$ \ $j$) | 0 | 1 | 2 | 3 | 4 | 5 | 6 | 7 | 8 |
|---|---|---|---|---|---|---|---|---|---|
| 0 | 0 | 0 | 0 | 0 | 0 | 0 | 0 | 0 | 0 |
| 1 | $W_2$ | 0 | $W_3$ | $W_3$ | $W_1$ | 0 | $W_3$ | 0 | 0 |
| 2 | 0 | $W_4$ | 0 | 0 | $W_4$ | 0 | $W_O$ | 0 | 0 |
| 3 | 0 | $W_4$ | 0 | 0 | 0 | 0 | $W_O$ | 0 | 0 |
| 4 | $W_2$ | $W_1$ | $W_3$ | 0 | $W_1$ | $W_3$ | $W_3$ | $W_3$ | $W_3$ |
| 5 | 0 | 0 | 0 | 0 | $W_4$ | 0 | 0 | $W_O$ | 0 |
| 6 | 0 | $W_4$ | $W_O$ | $W_O$ | $W_4$ | 0 | $W_O$ | $W_O$ | $W_O$ |
| 7 | 0 | 0 | 0 | 0 | $W_4$ | $W_O$ | $W_O$ | 0 | $W_O$ |
| 8 | 0 | 0 | 0 | 0 | $W_4$ | 0 | $W_O$ | $W_O$ | 0 |

*Table 2. Relationships between the vacancy jump frequency $W_{j\to k}$ from subset 'j' to subset 'k' and the standard notations of the five frequency model in a FCC lattice.*

### 3.3 Ingredients of the general formulation and result

In Appendix B, expressions (B1-B5) for the corrective terms appearing in the transport equation and corresponding to subsets $S_4$ to $S_8$ respectively are established. We get there the final formulation of the doubly transformed equation which fixes the expressions of the functions $g_i(k)$ as follows:

$$g_4(k) = -(W_2^{'} + 2W_1^{'} + 7W_3^{'})f_4 + W_3^{'}\left[4f_5 + f_6 + 2f_7 + f_8\right]$$

$$g_5(k) = \left[-4W_4^{'}f_5 + W_4^{'}f_4\right] \qquad g_6(k) = \left[-2W_4^{'}f_6 + 2W_4^{'}f_4\right]. \tag{28}$$

$$g_7(k) = \left[-2W_4^{'}f_7 + 2W_4^{'}f_4\right] \qquad g_8(k) = \left[-W_4^{'}f_8 + W_4^{'}f_4\right]$$

The system of equations is explicitly displayed and formally solved in Appendix B. When recast into its usual form, the correlation factor $f_{B*}$ for solute diffusion reads

$$f_{B*} = \frac{1+Q}{1-Q} = \frac{2W_1 + 7FW_3}{2W_2 + 2W_1 + 7FW_3},$$

where the function $F()$ is not expressed with the ratio $W_4 / W_0$ but with the reduced frequency $W_4^{'} = (W_4 - W_0) / W_0$.

The final value of the correlation factor obtained with this formulation agrees within 8 to 10 digits with the most accurate results previously published.[24]

## 4. General formulation for the case of solute-vacancy interaction of arbitrary range

As a result of the preceding section, it is clear that going beyond a first neighbour interaction becomes rapidly an unattractive task of tremendous complexity and widely open to errors. This is the reason why an automatic investigation of subset neighbourhoods was set up: the Fortran code in charge of this task is briefly described in Appendix C. In the latter it is explained how an automatic generation of the subroutine devoted to the calculation of lattice integrals avoids the building of a sparse matrix. The program then performs the numerical calculation of lattice integrals, solves the linear system yielding the desired return probabilities of the vacancy and evaluates the final value of the solute tracer correlation factor.

### 4.1 Universal notation for lattice integrals

For a reason detailed also in Appendix C, an alternative notation of functions $f_i$ or $F_i$ must be adopted in parallel. If a unique index $i$ is highly desirable for an automatic enumeration of subsets, its value is obviously linked to the order in which subsets are scanned and this order is changed when extending the range of the interaction. This is the reason why we will also name the function $f_i$ attached to subset $\left\{R_i^+\right\}$ with the help of the three coordinates $(i_1,i_2,i_3)$ of the *representative site* of the subset, i.e. $f_i \equiv f_{i_1 i_2 i_3}$ ; this notation points unambiguously towards the same function, whatever the number of the attached subset.

The coefficients $M_{ij}$ of Eqs. (27) are linear combinations of integrals $fifj(m,l) = -\frac{1}{n_m V_{ZB}} \int_{V_{ZB}} \frac{f_m(k) f_l(k)}{D_0} d_3k$ with the obvious identity $n_m fifj(m,l) = n_l fifj(l,m)$.
These integrals will be alternatively named

$$fifj_{m_1 m_2 m_3 \times l_1 l_2 l_3} = -\frac{1}{n_m V_{ZB}} \int_{V_{ZB}} \frac{f_{m_1 m_2 m_3}(k) f_{l_1 l_2 l_3}(k)}{D_0} d_3k \,,$$

and calculated once for all.

### 4.2 Defining a complete set of jump frequencies

The reference state of the system corresponds to the configuration where a solute tracer atom sits on the origin and the vacancy is far away in the bulk; the vacancy formation free energy is denoted $E_0 - TS_0$. When the vacancy is brought into a neighbour shell $j$ of the solute tracer atom, the free energy of the system becomes $E_0 + E_j - T(S_0 + S_j)$. The knowledge of the free energy changes $E_j - TS_j$ defines necessary conditions to be satisfied by the inward/outward jump frequencies on/from shell $j$ to shell $k$. We denote the frequencies by

$$W_{j\to k} = \upsilon_{j\to k} exp\left(-\frac{E^S_{j\leftrightarrow k} - E_j}{k_B T}\right) \qquad W_{k\to j} = \upsilon_{k\to j} exp\left(-\frac{E^S_{k\leftrightarrow j} - E_k}{k_B T}\right), \qquad (29)$$

where $k_B$ and $T$ stand for the Boltzmann constant and absolute temperature respectively, $E^S_{j\leftrightarrow k} = E^S_{k\leftrightarrow j}$ for the saddle energy with an atom at the saddle point between shell $j$ and $k$. To comply with thermodynamical consistency, the back and forth frequencies must obey the detailed balance

$$\frac{W_{j\to k}}{W_{k\to j}} = \frac{\upsilon_{j\to k}}{\upsilon_{k\to j}} exp\left(\frac{E_j - E_k}{k_B T}\right). \tag{30}$$

The ratio of the pre-exponential factors $\upsilon_{j\to k}$, $\upsilon_{k\to j}$ must obey

$$\frac{\upsilon_{j\to k}}{\upsilon_{k\to j}} = exp\left(\frac{S_k - S_j}{k_B}\right). \tag{31}$$

In the absence of any thorough ab initio evaluation of all the required quantities, further simplifications are made in practice:

* in most calculations, only the energy change which is linked to the formation energy in various shells is determined; the entropy change is ignored and the pre-exponential factors $\upsilon_{j\to k}$, $\upsilon_{k\to j}$ must then be taken as equal. Only the pre-exponential factor $\upsilon_{1\to 0}$ for the solute-vacancy exchange, which involves a foreign atom, is set to a different value;

* if detailed balancing imposes some constraints on the frequency ratio of direct and reverse jumps, it gives however no information on the way how to choose the absolute height of the saddles $E^S_{j\leftrightarrow k} = E^S_{k\leftrightarrow j}$ themselves.

To complete the missing quantities, two models are generally proposed:

* in model I, the energy of saddle configurations for jumps among the interactive sites is taken as a constant

$$E^S_{j\leftrightarrow k} = E^S_{k\leftrightarrow j} = E^S\,. \tag{32}$$

This implies that all the outward frequencies from a site with interaction energy $E_j$ towards any other site with $E_k$ have a migration barrier given by

$$E^M_{j\to k} = E^S - E_j \tag{33}$$

which depends only on the strength of the interaction at the departure site. When applied at large distances from the solute, model I requires for consistency $E^S = E^M_0$, where $E^M_0$ stands for the migration barrier in the bulk;

* in model II, the energy of the saddle configuration is given by

$$E^S_{j\leftrightarrow k} = E^S_{k\leftrightarrow j} = (E_j + E_k)/2 + \Delta\,. \tag{34}$$

As a result, the migration barrier from site with $E_j$ towards site with $E_k$ is given by

$$E^M_{j\to k} = \Delta + (E_k - E_j)/2\,, \tag{35}$$

which depends on the interaction energies at the departure and arrival sites. For the reverse jump, the migration barrier is given by $E^M_{k\to j} = \Delta + (E_j - E_k)/2$.

When applied at large distances from the solute, model II requires for consistency $\Delta = E^M_0$, where $E^M_0$ stands for the migration barrier in the bulk. This model is inspired from theoretical studies of Ising model for magnetism and numerical studies of diffusion.[26] But its relevance for vacancy jumps is still a pending question, since the underlying rule is followed neither in dilute alloys with ab-initio evaluations of saddle energies,[21,27-28] nor in homogeneous concentrated alloys with empirical potentials.[29]

### 4.3 Quantitative exploration of general models I and II

To explore quantitatively the implications of the above models, we chose an interaction, the absolute value of which decreases linearly with distance and vanishes between the 5th and 6th neighbour shell: $|E_1| > |E_2| > |E_3| > |E_4| > |E_5|$ with $|E_1| = 0.05$ and $|E_5| = 0.01$, the energy unit being the eV. The vacancy migration energy in the bulk is $E_0^M = 1$. The energy barrier for the solute-vacancy exchange $W_{10}$ is set to 0.9 eV. All the pre-exponential factors are set equal to $10^{13}$ $s^{-1}$. These values, although arbitrary, lie in a reasonable range for many metallic systems. The absolute temperature is arbitrarily fixed to 500K.

The interaction will be called attractive if $E_1 \leq 0$ and repulsive if $E_1 \geq 0$, regardless of the sign of the interaction at farther distances. To sweep all possible cases, the '+' and '-' signs will be used for the interaction energies beyond the first neighbour distance:

*case '+++++' stands for a monotonic repulsion at all distances and corresponds to $E_1 = +0.05,\ E_2 = +0.04,\ E_3 = +0.03,\ E_4 = +0.02,\ E_5 = +0.01$;

*case '+-+-+' stands for an oscillatory repulsion and corresponds to $E_1 = +0.05,\ E_2 = -0.04,\ E_3 = +0.03,\ E_4 = -0.02,\ E_5 = +0.01$;

*case '+- - - -' stands for a repulsion at first neighbour distance followed by an attraction at larger distances, i.e. $E_1 = +0.05,\ E_2 = -0.04,\ E_3 = -0.03,\ E_4 = -0.02,\ E_5 = -0.01$.

The repulsive/attractive case will then encompass 16 variants.

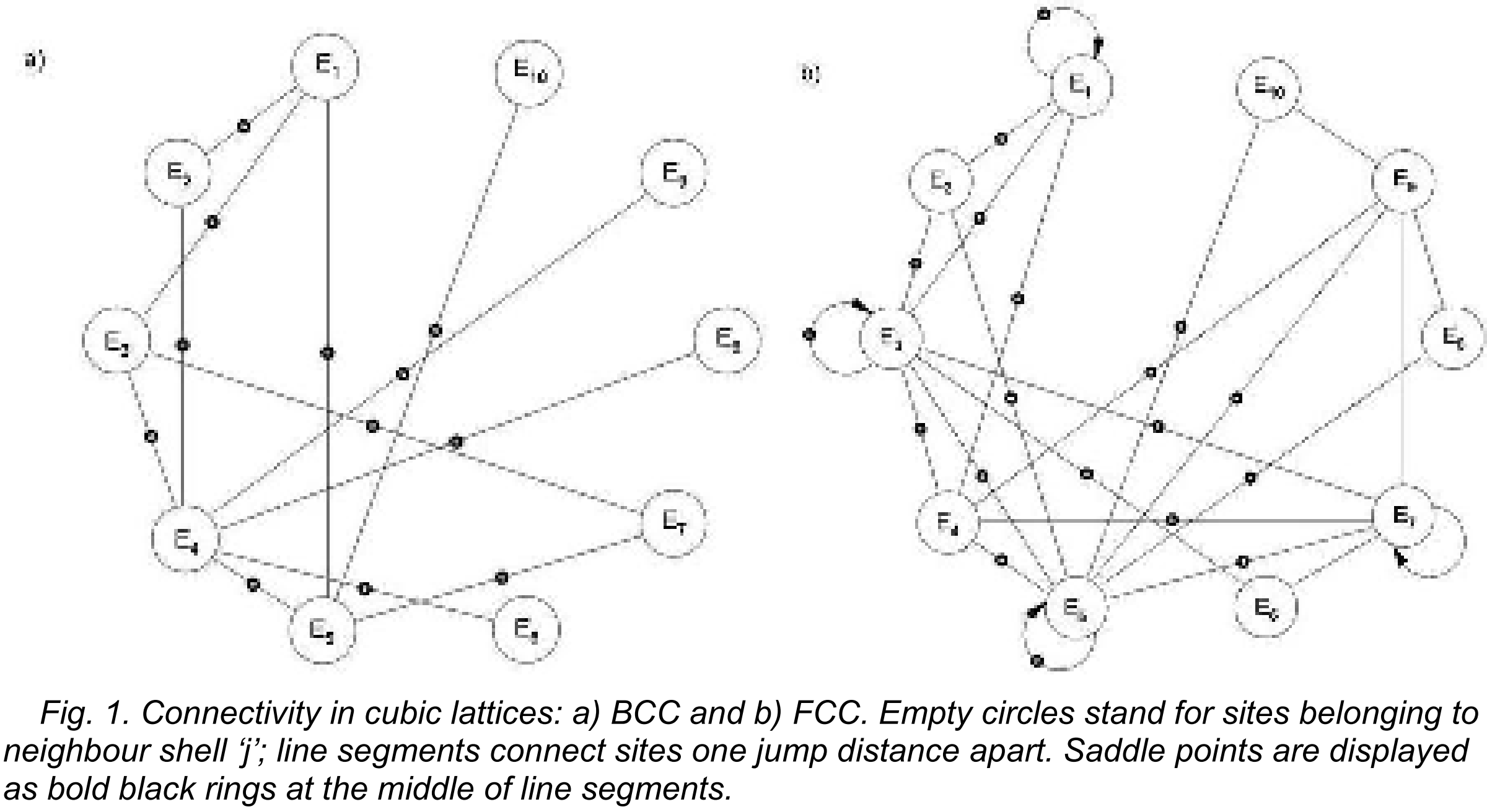

*Fig. 1. Connectivity in cubic lattices: a) BCC and b) FCC. Empty circles stand for sites belonging to neighbour shell 'j'; line segments connect sites one jump distance apart. Saddle points are displayed as bold black rings at the middle of line segments.*

The diagram of neighbourhoods displaying sites and saddles is shown for an interaction range extending up to 5th shell in BCC and FCC lattices (Fig. 1). The value enclosed in each circle is the solute-vacancy interaction energy $E_j$, which is set equal to zero if beyond the interaction radius. Arrows in closed loop correspond to jumps performed between sites belonging to the same shell, a situation which is encountered only in the FCC lattice (Fig. 1b). $E^S_{j\leftrightarrow k} = E^S_{k\leftrightarrow j}$ is the energy at the saddle point. After saddle energies have been calculated according to the models above, the associated migration barriers $E^M_{j\to k} = E^S_{j\leftrightarrow k} - E_j$ can be deduced and given as input data for the Fortran code. For a solute-vacancy interaction ranging up to 5th neighbour distance, the FCC lattice requires the knowledge of 19 saddle energies, versus only 12 for the BCC one. Each link or arrow, which is not decorated with a black ring, corresponds to a jump with a saddle energy equal to $E^M_0$ .

**4.3.1** Model I: $E^S_{j\leftrightarrow k} = E^S_{k\leftrightarrow j} = \Delta$

The results displayed on Fig. 2 are obtained with $\Delta$=1 eV. The striking fact is that the correlation factor is a constant (i.e. the results coincide within 12 digits) which does not depend on the range and on the signs which are retained for the interaction at 2nd to 5th neighbour distances.

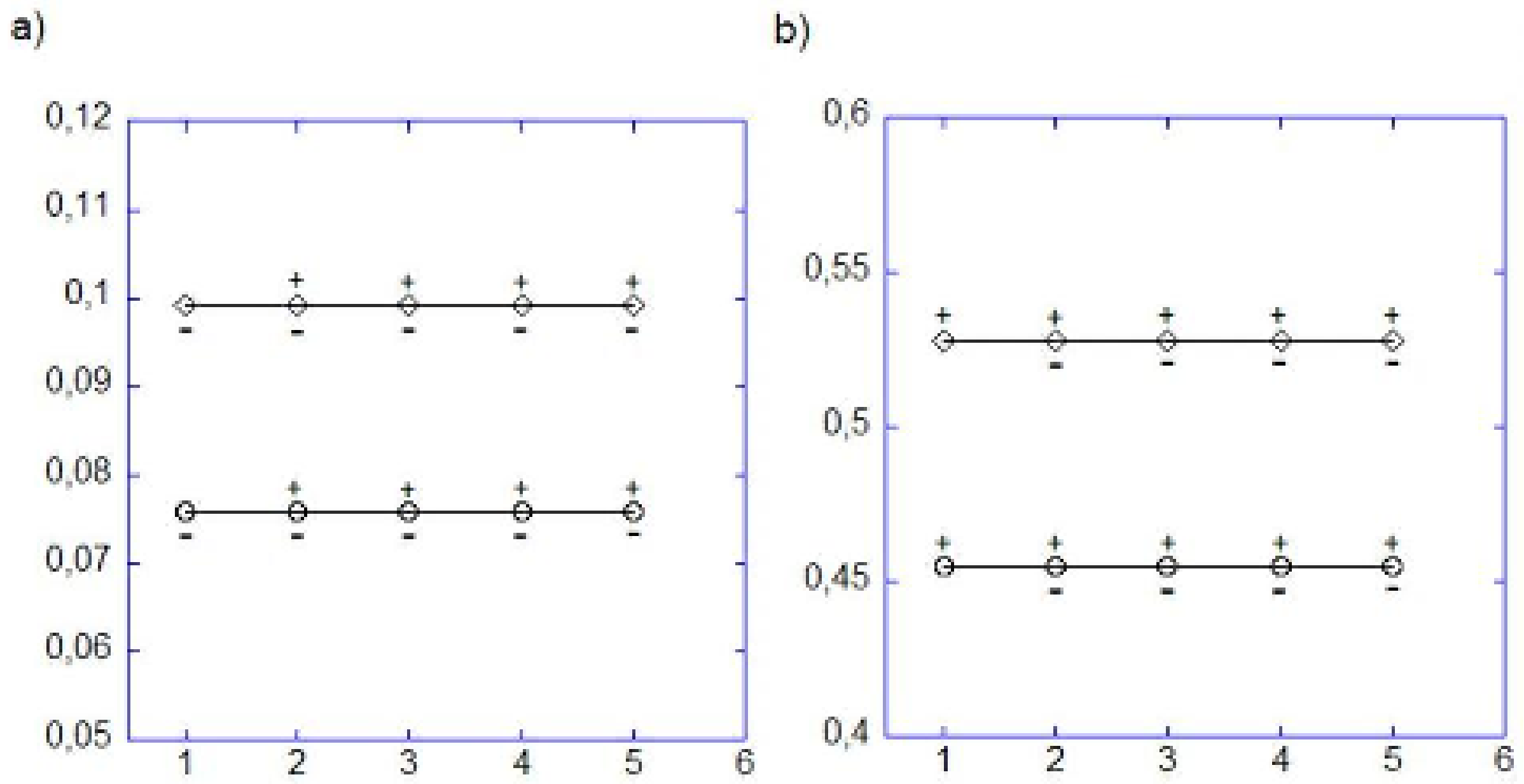

*Fig. 2. Solute correlation factor as a function of the interaction range in BCC (circles) and FCC (diamonds) lattices for solute-vacancy interactions of model I at 500K: a° attraction at first neighbour distance; b) repulsion at first neighbour distance.*

For the diagrams shown above in Fig. 1, the values of the energy barriers for the jump frequencies $W_{11}$, $W_{33}$, $W_{55}$, which link sites belonging to the same shell, were chosen in order to comply with the rule of model I, i.e. set equal to $E^S - E_1$, $E^S - E_3$, $E^S - E_5$ respectively. But the horizontal plateau is still observed if these migration barriers are chosen differently, the only difference being a shift along the ordinate axis. As a result, the final value of the correlation factor in model I depends only on the values of $E_1$, $W_{11}$,$W_{10}$ =$W_2$, $W_{33}$, $W_{55}$ but no longer on the others.

Although highly counter-intuitive, the reason can be traced back in the structure of the linear system in Eq. (27). The first equation (i.e. the one which monitors the occurrence of the second jump and fixes the value of the average product $<x_1x_2>$) links the return probability of the vacancy on the first neighbour shell to the return probabilities on farther sites. For model I, the cross-coefficients $M_{ij}$ (i≠j) of the first equation are all equal to zero and the coupling with the other unknowns of the problem disappears.

The case of an interaction up to the 5th shell in the FCC lattice is treated in more details in Appendix D. A close examination of all the ingredients entering the coefficient $M_{89}$ suggests the identity $12 fifj(8,9) = fifj(8,8) + fifj(8,11) + fifj(8,14)$ which can be written as well under the form $12 fifj_{101\times200} = fifj_{101\times101} + fifj_{101\times211} + fifj_{101\times301}$. This is easy to establish formally (Appendix D). This identity is supplemented by three other ones, all of them being displayed in Table 3 below. It is obvious that other relationships do exist, but they will modify only the links between more distant return probabilities and will not alter further the correlation factor.

The same work can be done for the BCC lattice; in the latter, five identities are found because the 1rst neighbour shell is connected though a jump with the 5th neighbour shell. The relationships are gathered in Table 3.
We keep mentioning the structure of the underlying lattice to avoid a misleading interpretation of the notations. $F_{ijk}$ has an analytical expression which depends only on indices 'I', 'j' and 'k' and is obviousy independent of the lattice structure (for instance $F_{200}$ which is used in both lattices). However the integral embodied in the notation $fifj_{ijk\times lmn}$ has a denominator $D_0$ which does depend on the lattice structure according to Eqs. (16-17) above.

| |
|---|
| FCC |
| $12 fifj_{101\times200} = fifj_{101\times101} + fifj_{101\times211} + fifj_{101\times301}$ |
| $11 fifj_{101\times112} = 2 fifj_{101\times101} + fifj_{101\times211} + fifj_{101\times202} + fifj_{101\times103} + fifj_{101\times222}$ |
| $12 fifj_{101\times211} = 2 fifj_{101\times101} + 4 fifj_{101\times200} + fifj_{101\times112} + 2 fifj_{101\times202} + 2 fifj_{101\times301} + fifjj_{101\times222} + fifj_{101\times312}$ |
| $12 fifj_{101\times202} = fifj_{101\times101} + fifj_{101\times112} + 2 fifj_{101\times211} + fifj_{101\times103} + fifj_{101\times301} + fifj_{101\times213} + fifj_{101\times312} + fifj_{101\times303}$ |
| BCC |
| $8 fifj_{111\times200} = fifj_{111\times111} + fifj_{111\times311}$ |
| $8 fifj_{111\times202} = 2 fifj_{111\times111} + fifj_{111\times113} + 2 fifj_{111\times311} + fifj_{111\times313}$ |
| $8 fifj_{111\times113} = 2 fifj_{111\times202} + fifj_{111\times222} + 2 fifj_{111\times204} + fifj_{111\times224}$ |
| $8 fifj_{111\times311} = 4 fifj_{111\times200} + 2 fifj_{111\times202} + fifj_{111\times222} + 4 fifj_{111\times400} + 2 fifj_{111\times402} + fifj_{111\times422}$ |
| $8 fifj_{111\times222} = fifj_{111\times111} + fifj_{111\times113} + fifj_{111\times311} + fifj_{111\times133} + fifj_{111\times313} + fifj_{111\times333}$ |

*Table 3. Relationships between lattice integrals entering the return probabilities*

**4.3.2** Model II: $E^S_{j\leftrightarrow k} = E^S_{k\leftrightarrow j} = (E_j + E_k)/2 + \Delta$

As explained above, $\Delta$ was set to 1 eV for consistency with the vacancy migration energy $E_0^M$ far from the solute tracer atom. The results are gathered in the diagrams of Fig. 3 for BCC and Fig. 4 for FCC lattices: the lines were drawn to connect properly the sequence of data which are obtained when extending the range of the interaction.

Several general trends are clearly displayed:

* the range encompassing the final values depends primarily on the sign of the interaction energy at 1rst neighbour site: attractive cases (left diagrams) yield lower values than repulsive ones (right diagrams);

* in all cases (attractive or repulsive), choosing the interactions at farther distances to be all attractive yields larger correlation factors than choosing them all repulsive. The upper envelope corresponds to the case '± - - - -' and the lower one to '± + + + +';

* mixing attractive and repulsive interactions ends up with a correlation factor lying always in between these envelopes: the sooner the sign change, the larger the compensation between the repulsive and attractive interactions. The oscillating cases '- + - + -' or '+ - + -+' provide final results which are nearly equal to the result obtained with an interaction at first neighbour only;

* at each step, adding a further interaction with the '+' or '-' choice always produces a splitting of the subsequent results, that obtained with the '-' sign being always larger than the one obtained with the '+' sign. Beyond the third neighbour shell, the final hierarchy of results displays always the alternation '+-+- ….' from bottom to top. The only exception is the cross-over of the curves obtained in the two structures when extending the interaction from 2nd neighbour distance to 3rd neighbour distance.

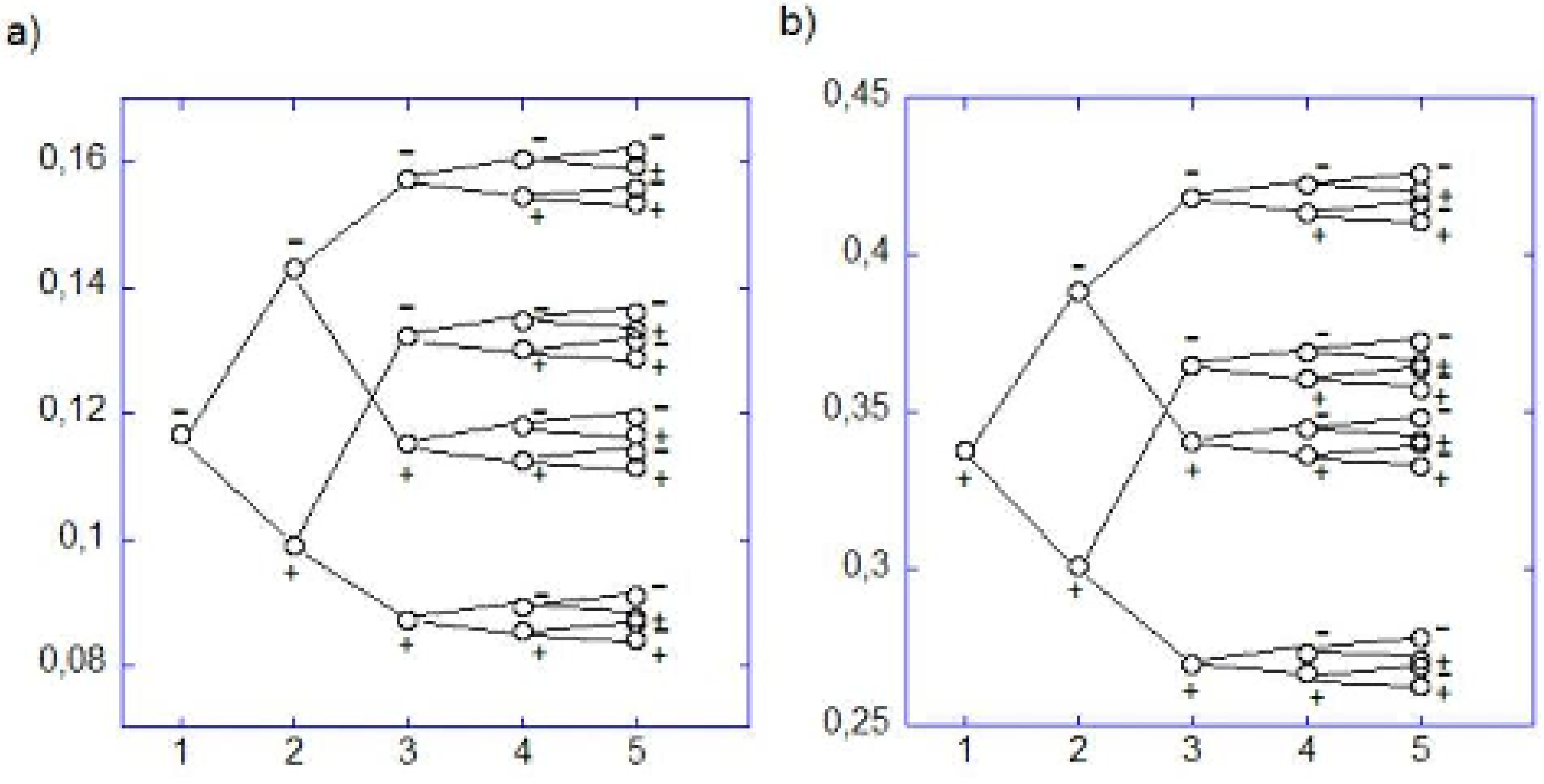

*Fig. 3. Solute correlation factor as a function of the interaction range in a BCC lattice for solute-vacancy interactions of model II at 500K: a) attraction at first neighbour distance; b) repulsion at first neighbour distance. Lines are drawn to guide the eye along the sequence of interaction signs.*

The general conclusion is that a thorough knowledge of interaction energies and migration barriers up to 3rd neighbours at least is mandatory to know the correlation factor and, as a consequence, the solute tracer diffusion coefficient with a reasonable accuracy. The reader should remember that this conclusion was obtained with interaction energies which decrease with distance, a feature which is not always exhibited in actual systems.

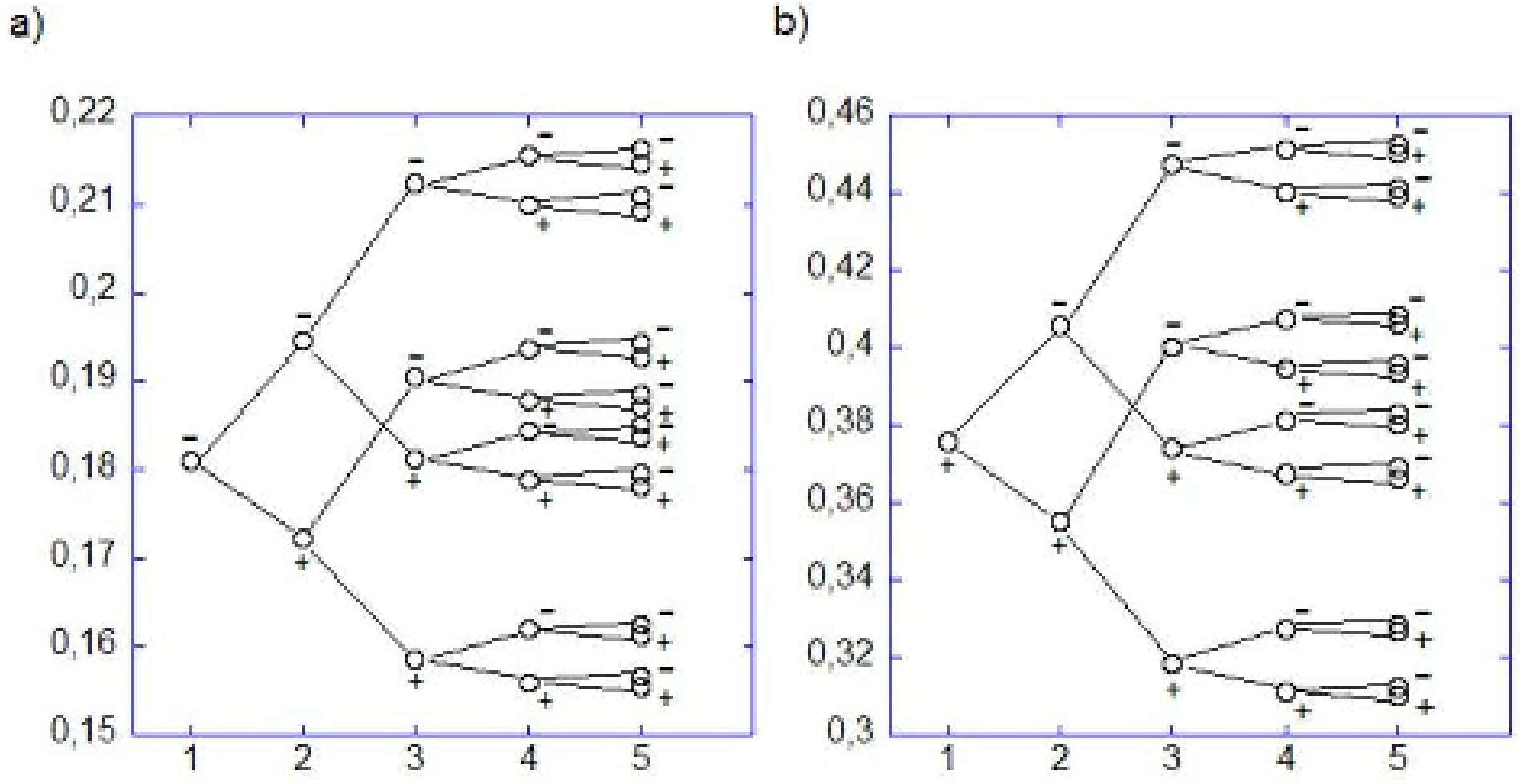

*Fig. 4. Solute correlation factor as a function of the interaction range in a FCC lattice for solute-vacancy interactions of model II at 500K: a) attraction at first neighbour distance; b) repulsion at first neighbour distance. Lines are drawn to guide the eye along the sequence of interaction signs.*

## 5. Examination of several results formerly published in the literature

### ***Ni-Cr alloys*** [21]

Let us notice first that the migration barriers of Table D.2 [21] are not consistent with the solute vacancy interactions displayed on Fig. D.1. In Table 4 below, the first column reports the interaction energies which can be read on the diagram (set to zero beyond the 5$^{th}$ neighbour shell); the second column reports the migration barriers from/to a first neighbour site; the third column contains the corresponding height of the saddle points which can be deduced by adding the two first columns. Saddle energies should be equal for direct and reverse jumps, i.e. $E^S_{j\leftrightarrow k} = E^S_{k\leftrightarrow j}$, a feature which is not observed. Migration barriers are less precisely evaluated than energies in stable positions and the authors estimate the error to be less than 0.035 eV. This is reason why we kept the original values of interaction energies and corrected for consistency only the saddle energies by quantities which never exceeded ±0.02 eV. The corrected values are written in red in the fourth and fifth columns.

Values for bulk and solute-vacancy migration barriers, together with the values of pre-exponential factors are kept to the original values:
$E^M_{Ni} = 1.090\ eV$ $\upsilon_{Ni} = 4.48\ 10^{12} s^{-1}$ and $E^M_{Cr} = 0.89\ eV$ $\upsilon_{Cr} = 4.92\ 10^{12} s^{-1}$.

The missing saddle energies must then be fixed: they are calculated with model I or II. All other frequencies are assumed to be equal to the Ni self-diffusion frequency.

The diagrams of Figs. 5-6 display the ingredients giving access to the migration barriers.

| Interaction energy (eV) | Original barrier (eV) | Original saddle (eV) | Corrected saddle (eV) | Corrected barrier (eV) |
|---|---|---|---|---|
| $E_1 = +0.045$ | $E^M_{11} = 0.98$<br>$E^M_{12} = 1.02$<br>$E^M_{13} = 1.04$<br>$E^M_{14} = 1.04$ | $E^S_{11} = 1.025$<br>$E^S_{12} = 1.065$<br>$E^S_{13} = 1.085$<br>$E^S_{14} = 1.085$ | $E^S_{11} = 1.025$<br>$E^S_{12} = 1.064$<br>$E^S_{13} = 1.090$<br>$E^S_{14} = 1.090$ | $E^M_{11} = 0.98$<br>$E^M_{12} = 1.019$<br>$E^M_{13} = 1.045$<br>$E^M_{12} = 1.045$ |
| $E_2 = +0.013$ | $E^M_{21} = 1.05$ | $E^S_{21} = 1.063$ | $E^S_{21} = 1.064$ | $E^M_{21} = 1.051$ |
| $E_3 = +0.036$ | $E^M_{31} = 1.06$ | $E^S_{31} = 1.096$ | $E^S_{31} = 1.090$ | $E^M_{21} = 1.054$ |
| $E_4 = -0.036$ | $E^M_{41} = 1.14$ | $E^S_{41} = 1.104$ | $E^S_{41} = 1.090$ | $E^M_{21} = 1.126$ |
| $E_5 = +0.011$ | | | | |

*Table 4. Interaction energies and original values for migration barriers by Tucker in Ni(Cr) alloys; retained values for saddles and migration barriers chosen for consistency in 4th and 5th columns.*

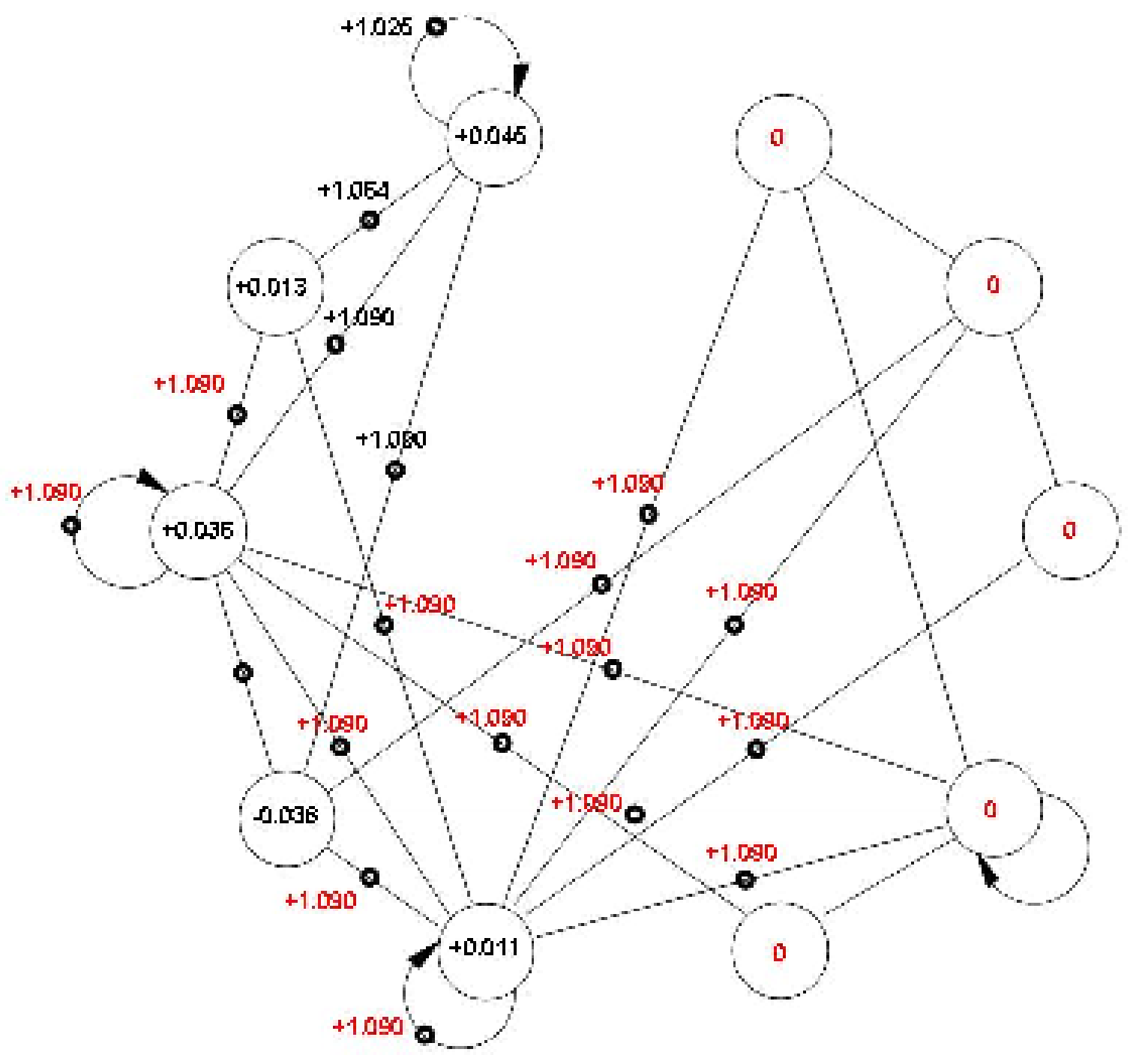

*Fig. 5. Diagram of saddle energies for Ni(Cr) alloys for the full interaction range up to the 5th shell ; original values by Tucker are written in black; values completed with model I are written in red. Links without a black ring correspond to a saddle energy equal to the migration energy in the bulk.*

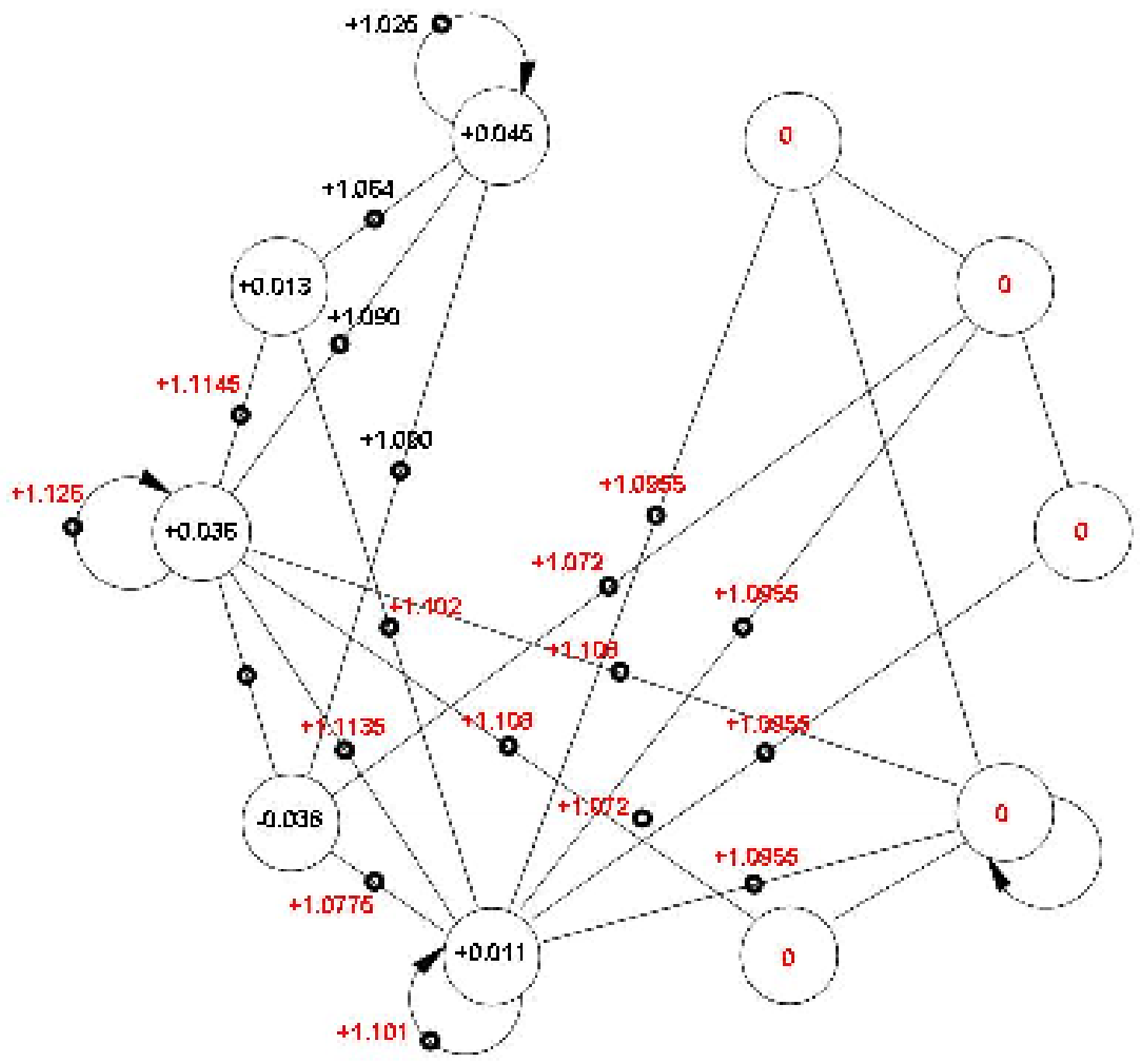

*Fig. 6. Diagram of saddle energies for Ni(Cr) alloys for the full interaction range up to the 5^th shell; original values by Tucker are written in black; values completed with model II are written in red.*

Reducing the range of the interaction is performed in the frame of the two models as follows:

* for model I (Fig. 5), the reduction from 5th to 4th neighbour interaction is carried out through setting $E_5 = 0$. As a consequence, the migration barriers for all the outgoing frequencies from a site lying in the 5th shell become equal to 1.09 eV instead of 1.09 eV - 0.011 eV …etc;

* for model II (Fig. 6), setting $E_5 = 0$ modifies the saddle energies $E^S_{5\leftrightarrow i}$ for i=2,3,4,5,7,8,9,10. As above, the corresponding migrations barriers for the outgoing frequencies from a site lying in the 5th shell become equal to the corresponding new saddle energies $E^M_{5\to i} = E^S_{5\leftrightarrow i}$.

The results are displayed on Fig. 7a. All the trends observed with the general models are present:

* a constant value for model I;

* oscillations in agreement with the sign of the next shell included in the interaction for model II, namely the sharp drop when adding the repulsive interaction at 3rd neighbour followed by the rise due to the addition of an attractive 4th neighbour interaction.

The value obtained with the migration barriers retained by Tucker for the five-frequency model, and which is inconsistent as explained above, namely $E^M_{12} = 1.02$ ; $E^M_{13} = E^M_{14} = 1.04$ ; $E^M_{21} = 1.51$ ; $E^M_{31} = E^M_{41} = 1.06$ is also plotted for comparison.

The overall change is of the order of 8%, i.e. not very large but noticeable.

***Ni-Fe alloys*** [21]

The same observations as above can be made on the inconsistency of the published data for migration barriers. The corrected values are recalled below in Table 5. Values for bulk and solute-vacancy migration barriers, together with the values of pre-exponential factors are kept to the original values:
$E^M_{Ni} = 1.090\ eV$ $\quad \upsilon_{Ni} = 4.48\ 10^{12} s^{-1}$, and $E^M_{Fe} = 0.97\ eV$ $\quad \upsilon_{Fe} = 4.14\ 10^{12} s^{-1}$.

| Interaction energy (eV) | Original barrier (eV) | Original saddle (eV) | Corrected saddle (eV) | Corrected barrier (eV) |
|---|---|---|---|---|
| $E_1 = +0.025$ | $E^M_{11} = 1.13$<br>$E^M_{12} = 1.07$<br>$E^M_{13} = 1.08$<br>$E^M_{14} = 1.05$ | $E^S_{11} = 1.155$<br>$E^S_{12} = 1.095$<br>$E^S_{13} = 1.105$<br>$E^S_{14} = 1.075$ | $E^S_{11} = 1.155$<br>$E^S_{12} = 1.090$<br>$E^S_{13} = 1.100$<br>$E^S_{14} = 1.085$ | $E^M_{11} = 1.13$<br>$E^M_{12} = 1.065$<br>$E^M_{13} = 1.075$<br>$E^M_{14} = 1.060$ |
| $E_2 = +0.020$ | $E^M_{21} = 1.07$ | $E^S_{21} = 1.090$ | $E^S_{21} = 1.090$ | $E^M_{21} = 1.070$ |
| $E_3 = +0.032$ | $E^M_{31} = 1.07$ | $E^S_{31} = 1.102$ | $E^S_{31} = 1.100$ | $E^M_{31} = 1.068$ |
| $E_4 = -0.016$ | $E^M_{41} = 1.11$ | $E^S_{41} = 1.094$ | $E^S_{41} = 1.085$ | $E^M_{41} = 1.101$ |
| $E_5 = +0.012$ | | | | |

*Table 5. Interaction energies and original values for migration barriers by Tucker in Ni(Fe) alloys; retained values for saddles and migration barriers chosen for consistency in 4$^{th}$ and 5$^{th}$ columns.*

The results are displayed on Fig. 7b and the same trends as above are observed. The value obtained with the (inconsistent) choice of migration barriers by Tucker for the five-frequency model, namely $E^M_{12} = E^M_{13} = E^M_{14} = 1.07$ ; $E^M_{21} = E^M_{31} = E^M_{41} = 1.07$ is also plotted for comparison.
The overall change is of the order of 10%, i.e. not very large but noticeable.

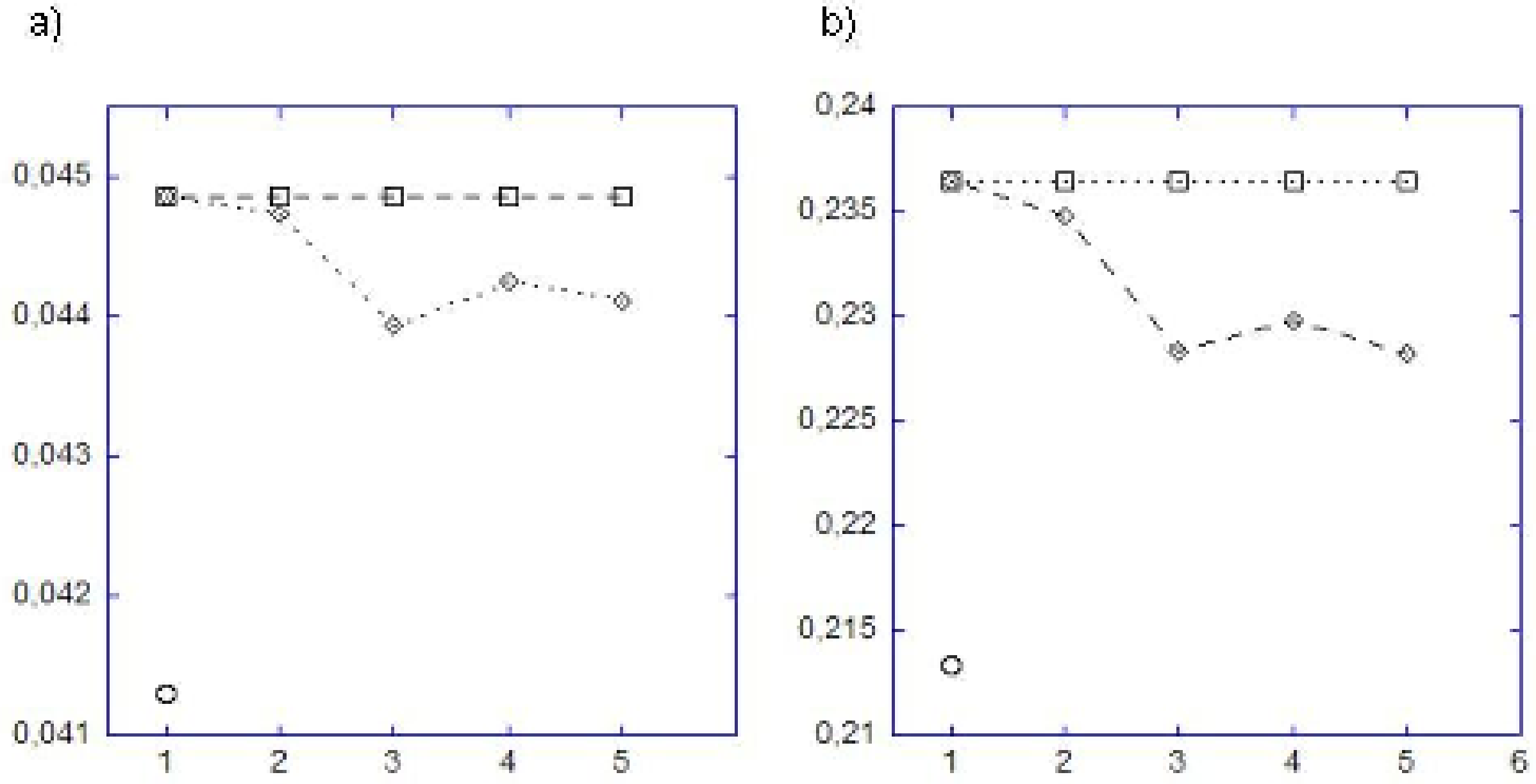

*Fig. 7. Tracer correlation factor in Ni at 500K for model I (squares) and II (diamonds) as a function of the solute-vacancy interaction range retained in the calculation: a) Cr* tracer ; b) Fe* tracer. The original value by Tücker is displayed as an empty circle..*

### ***Ni-Si alloys*** [28]

Table I of the paper is the first ever published which contains nearly all the ingredients required to determine a fully consistent pattern of vacancy jump frequencies around a solute atom of Si in Ni.[28] The vacancy-silicon interaction is determined up to the 3rd neighbour distance and set to zero beyond. Since a 3rd neighbour site is first neighbour of a 7th neighbour site, a few saddle energies are however still lacking, namely those controlling the frequencies $W_{35}$, $W_{36}$, $W_{37}$ and $W_{45}$, $W_{47}$, $W_{55}$, $W_{57}$, $W_{77}$. The saddle energies for the frequencies belonging to the second set correspond to jumps which take place entirely outside the interaction sphere; they can be reasonable taken equal to the saddle energy in the pure bulk $E_0^M$, i.e. 1.074 eV. The saddle energies for the remaining escape frequencies belonging to the first set $W_{35}$, $W_{36}$, $W_{37}$ can be equated either to the saddle energy of jump $W_{34}$, i.e. equal to 1.113 eV, or to the saddle energies in the pure bulk $E_0^M$, i.e.1.074 eV (Fig. 8). The retained values for the pre-exponential factors are

$\upsilon_{1\to 0} = \upsilon_{Si} = 5.10^{12} s^{-1}$ $\qquad$ $\upsilon_{0\to 0} = \upsilon_{Ni} = 5.10^{12} s^{-1}$.

The correlation factor for Si* tracer at 1000 K changes only from 0.26906 in the first case to 0.27106 in the second case.

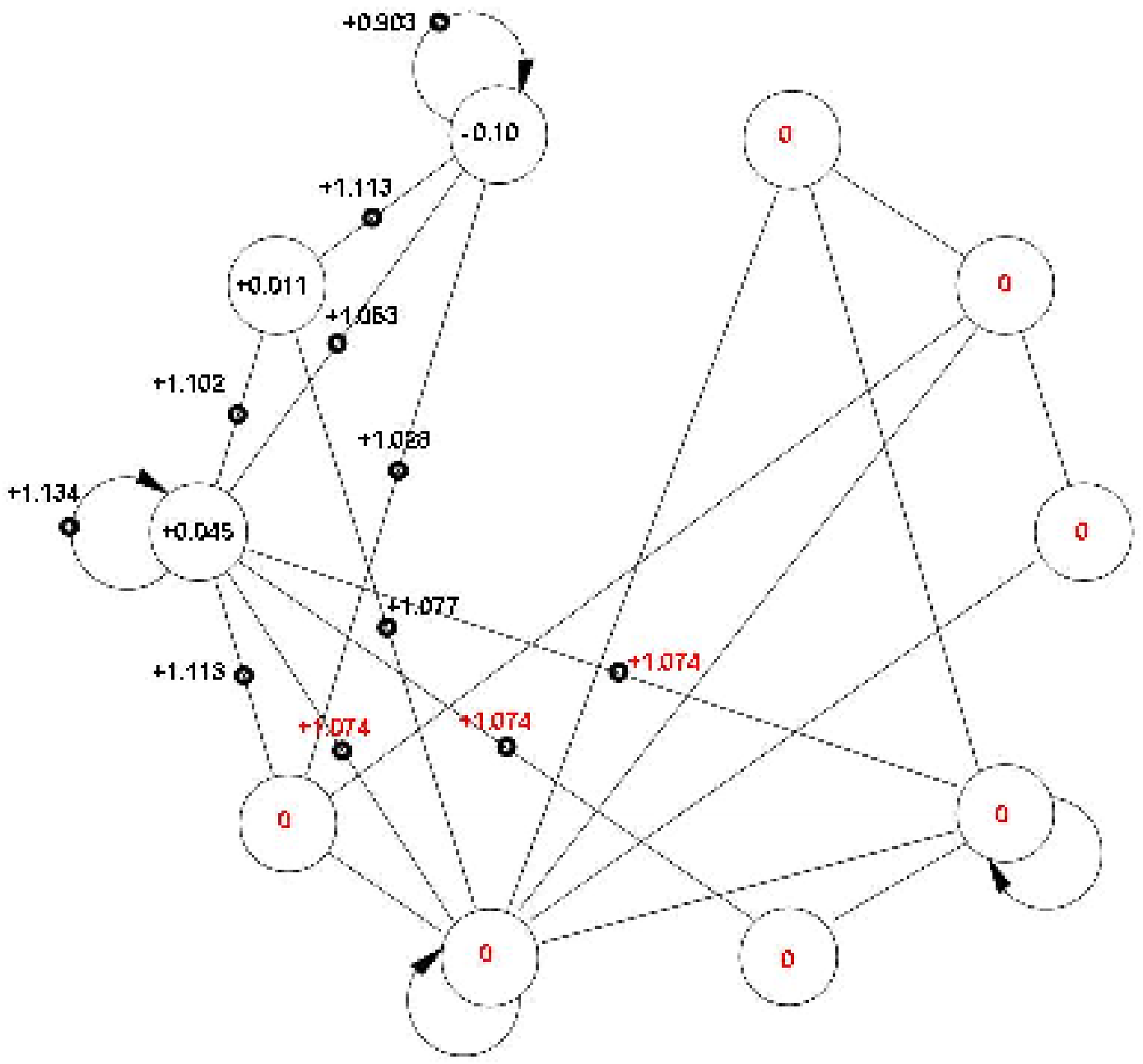

*Fig.8. Diagram of site and saddle energies for Ni(Si) alloys; saddle energies written in black are published values; saddle energies written in red are added to complete the modelling. All other saddle energies are set to 1.074 eV.*

We can also evaluate, in the frame of model I and II, the correlation factor for Si* tracer in Ni, *if only the interaction energies were known*, while keeping $E_{11}^S$ and $E_{10}^S$ equal to the published value of 0.903 eV and 0.891 eV respectively. The saddle energies which were written in black in Fig. 8 are then modified as follows :

*in model I, all saddle energies are set to 1.074 eV: the correlation factor amounts to 0.26732 at 1000K;

*in model II, with saddle energies evaluated according to Eq. (34) with Δ=1.074 eV, the correlation factor amounts to 0.28384.

At last, taking into account only the interaction at 1st neighbour distance as done in the five-frequency model, the dissociative frequencies $W_{1\to2}$, $W_{1\to3}$, $W_{1\to4}$ and reassociative frequencies $W_{2\to1}$, $W_{3\to1}$, $W_{4\to1}$ are replaced respectively by an average frequency $<W_3>=2W_{1\to2}+4W_{1\to3}+W_{1\to4}$ and $<W_4>=2W_{2\to1}+4W_{3\to1}+W_{4\to1}$ calculated at 1000K. The correlation factor amounts to 0.26858.

In the case of Ni(Si) alloys, the overall change is detectable but remains small, thanks to the fact that nearly all the necessary frequencies were previously determined.

**Fe-Cu alloys** [22]

The original ab-initio results for interaction and migration energies are displayed in the first two columns of Tables 6-7. Interaction energies are not given beyond the 2nd neighbour shell. Two different ways of exploiting these data are explored hereafter:

* starting from the published interaction and migration energies, saddle energies can be deduced, which have then to be slightly corrected for consistency. Namely $E^S_{1\leftrightarrow2}$, $E^S_{1\leftrightarrow3}$, $E^S_{1\leftrightarrow5}$, $E^S_{2\leftrightarrow4}$ are averaged to 0.507, 0.566, 0.518 and 0.576 eV respectively, which modifies slightly the corresponding migration energies. The corrected values of saddle and migration energies are reported in red in columns 4 and 5 of Table 6;

* conversely original migration barriers and saddle energies can be kept as is (or nearly so, apart from the necessary correction for $E^S_{1\leftrightarrow2}$), but additional interaction energies at 3rd, 4th and 5th neighbour shells are introduced in order to comply fully with the values of the migration and saddle energies, namely, $E_3=+0.013eV$, $E_4=+0.052eV$, $E_5=-0.027eV$, which are slightly repulsive but the last (Table 7).

For all quantities which are not calculated in the original paper, model II is used throughout with $\Delta=E^M_0=0.68\ eV$. Pre-exponential factors and migration energies for Fe and Cu are kept as is:

$E^M_{Fe}=E^M_0=0.68\ eV \quad \upsilon_{Fe}=5\ 10^{15}s^{-1}$ and $E^M_{Cu}=0.59\ eV \quad \upsilon_{Cu}=2\ 10^{15}s^{-1}$.

Reducing the range of the interaction from 5th to 4th, 3rd and 2nd neighbour is performed by successively setting the corresponding interaction energies to zero and restoring progressively the saddle energies $E^S_{5\leftrightarrow1}$, $E^S_{4\leftrightarrow2}$, $E^S_{3\leftrightarrow1}$ to their average values i.e. 0.518, 0.576 and 0.566 eV respectively.

In agreement with the trends highlighted in the general models, the positive 3rd and 4th neighbour interaction decrease the correlation factor from 0.5188 down to 0.5111; the 5th neighbour interaction is negative and increases the result up to 0.5127.

| Interaction energies | Original migration barriers | Original saddle energies | Corrected saddle energies | Corrected migration barriers |
|---|---|---|---|---|
| $E_1 = -0.127$ | $E^M_{1\to2} = 0.64$<br>$E^M_{1\to3} = 0.70$<br>$E^M_{1\to5} = 0.63$ | $E^S_{1\leftrightarrow2} = 0.513$<br>$E^S_{1\leftrightarrow3} = 0.573$<br>$E^S_{1\leftrightarrow5} = 0.503$ | $E^S_{1\leftrightarrow2} = 0.507$<br>$E^S_{1\leftrightarrow3} = 0.566$<br>$E^S_{1\leftrightarrow5} = 0.518$ | $E^M_{1\to2} = 0.634$<br>$E^M_{1\to3} = 0.693$<br>$E^M_{1\to5} = 0.645$ |
| $E_2 = -0.138$ | $E^M_{2\to1} = 0.64$<br>$E^M_{2\to4} = 0.74$ | $E^S_{2\leftrightarrow1} = 0.502$<br>$E^S_{2\leftrightarrow4} = 0.602$ | $E^S_{2\leftrightarrow1} = 0.507$<br>$E^S_{2\leftrightarrow4} = 0.576$ | $E^M_{2\to1} = 0.645$<br>$E^M_{2\to4} = 0.714$ |
| $E_3 = 0$ | $E^M_{3\to1} = 0.56$<br>$E^M_{3\to4} = 0.68$<br>$E^M_{3\to7} = ?$ | $E^S_{3\leftrightarrow1} = 0.56$<br>$E^S_{3\leftrightarrow4} = 0.68$<br>$E^S_{3\leftrightarrow7} = ?$ | $E^S_{3\leftrightarrow1} = 0.566$<br>$E^S_{3\leftrightarrow4} = 0.68$<br>$E^S_{3\leftrightarrow7} = 0.68$ | $E^M_{3\to1} = 0.566$<br>$E^M_{3\to4} = 0.68$<br>$E^M_{3\to7} = 0.68$ |
| $E_4 = 0$ | $E^M_{4\to2} = 0.55$<br>$E^M_{4\to3} = 0.68$<br>$E^M_{4\to5} = ?$<br>$E^M_{4\to6} = ?$<br>$E^M_{4\to8} = ?$<br>$E^M_{4\to9} = ?$ | $E^S_{4\to2} = 0.55$<br>$E^S_{4\to3} = 0.68$<br>$E^S_{4\to5} = ?$<br>$E^S_{4\to6} = ?$<br>$E^S_{4\to8} = ?$<br>$E^S_{4\to9} = ?$ | $E^S_{4\to2} = 0.576$<br>$E^S_{4\to3} = 0.68$<br>$E^S_{4\to5} = 0.68$<br>$E^S_{4\to6} = 0.68$<br>$E^S_{4\to8} = 0.68$<br>$E^S_{4\to9} = 0.68$ | $E^M_{4\to2} = 0.576$<br>$E^M_{4\to3} = 0.68$<br>$E^M_{4\to5} = 0.68$<br>$E^M_{4\to6} = 0.68$<br>$E^M_{4\to8} = 0.68$<br>$E^M_{4\to9} = 0.68$ |
| $E_5 = 0$ | $E^M_{5\to1} = 0.53$<br>$E^M_{5\to4} = ?$<br>$E^M_{5\to7} = ?$<br>$E^M_{5\to10} = ?$ | $E^S_{5\leftrightarrow1} = 0.53$<br>$E^S_{5\leftrightarrow4} = ?$<br>$E^S_{5\leftrightarrow7} = ?$<br>$E^S_{5\leftrightarrow10} = ?$ | $E^S_{5\leftrightarrow1} = 0.518$<br>$E^S_{5\leftrightarrow4} = 0.68$<br>$E^S_{5\leftrightarrow7} = 0.68$<br>$E^S_{5\leftrightarrow10} = 0.68$ | $E^M_{5\to1} = 0.518$<br>$E^M_{5\to4} = 0.68$<br>$E^M_{5\to7} = 0.68$<br>$E^M_{5\to10} = 0.68$ |

*Table 6. Interaction energies and original values of migration barriers for Fe(Cu) alloys are displayed in columns 1 and 2; deduced saddle energies in column 3; corrected values of saddle and migration energies are written in red in columns 4 and 5.*

At last the value obtained in the frame of a five-frequency model, although not perfectly suited to the BCC lattice, was also calculated for comparison. Taking into account only the interaction at 1st neighbour distance, the dissociative frequencies $W_{1\to2}$, $W_{1\to3}$, $W_{1\to5}$ and reassociative frequencies $W_{2\to1}$, $W_{3\to1}$, $W_{5\to1}$ are replaced respectively by the average frequency $<W_3>= 3W_{1\to2} + 3W_{1\to3} + W_{1\to5}$ and $<W_4>= 3W_{2\to1} + 3W_{3\to1} + W_{5\to1}$ calculated at 500K. The correlation factor drops down to 0.49633.

For this alloy, the overall change remains small of the order of 4%.

| Original interaction energies | Original migration barriers | Original saddle energies | Corrected interaction energies | Retained migration barriers |
|---|---|---|---|---|
| $E_1 = -0.127$ | $E^M_{1\to 2} = 0.64$<br>$E^M_{1\to 3} = 0.70$<br>$E^M_{1\to 5} = 0.63$ | $E^S_{1\leftrightarrow 2} = 0.507$<br>$E^S_{1\leftrightarrow 3} = 0.573$<br>$E^S_{1\leftrightarrow 5} = 0.503$ | | $E^M_{1\to 2} = 0.634$<br>$E^M_{1\to 3} = 0.70$<br>$E^M_{1\to 5} = 0.63$ |
| $E_2 = -0.138$ | $E^M_{2\to 1} = 0.64$<br>$E^M_{2\to 4} = 0.74$ | $E^S_{2\leftrightarrow 1} = 0.507$<br>$E^S_{2\leftrightarrow 4} = 0.602$ | | $E^M_{2\to 1} = 0.645$<br>$E^M_{2\to 4} = 0.74$ |
| $E_3 = 0$ | $E^M_{3\to 1} = 0.56$<br>$E^M_{3\to 4} = 0.68$<br>$E^M_{3\to 7} = ?$ | $E^S_{3\leftrightarrow 1} = 0.573$<br>$E^S_{3\leftrightarrow 4} = 0.68$<br>$E^S_{3\leftrightarrow 7} = ?$ | $E_3 = +0.013$ | $E^M_{3\leftrightarrow 1} = 0.56$<br>$E^M_{3\leftrightarrow 4} = 0.6995$<br>$E^M_{3\leftrightarrow 7} = 0.6735$ |
| $E_4 = 0$ | $E^M_{4\to 2} = 0.55$<br>$E^M_{4\to 3} = 0.68$<br>$E^M_{4\to 5} = ?$<br>$E^M_{4\to 6} = ?$<br>$E^M_{4\to 8} = ?$<br>$E^M_{4\to 9} = ?$ | $E^S_{4\to 2} = 0.602$<br>$E^S_{4\to 3} = 0.68$<br>$E^S_{4\to 5} = ?$<br>$E^S_{4\to 6} = ?$<br>$E^S_{4\to 8} = ?$<br>$E^S_{4\to 9} = ?$ | $E_4 = +0.052$ | $E^M_{4\to 2} = 0.55$<br>$E^M_{4\to 3} = 0.6605$<br>$E^M_{4\to 5} = 0.6405$<br>$E^M_{4\to 6} = 0.654$<br>$E^M_{4\to 8} = 0.654$<br>$E^M_{4\to 9} = 0.654$ |
| $E_5 = 0$ | $E^M_{5\to 1} = 0.53$<br>$E^M_{5\to 4} = ?$<br>$E^M_{5\to 7} = ?$<br>$E^M_{5\to 10} = ?$ | $E^S_{5\leftrightarrow 1} = 0.503$<br>$E^S_{5\leftrightarrow 4} = ?$<br>$E^S_{5\leftrightarrow 7} = ?$<br>$E^S_{5\leftrightarrow 10} = ?$ | $E_5 = -0.027$ | $E^M_{5\to 1} = 0.530$<br>$E^M_{5\to 4} = 0.7195$<br>$E^M_{5\to 7} = 0.6935$<br>$E^M_{5\to 10} = 0.6935$ |

*Table 7. Interaction energies and original values of migration barriers for Fe(Cu) alloys are displayed in columns 1 and 2; deduced saddle energies in column 3; corrected values of interaction energies are written in red in columns 4. Retained migration barriers are reported in column 5.*

## Conclusion

A formalism using a double Laplace and Fourier transform of the transport equation for the vacancy yields the return probabilities of the vacancy in the close vicinity of the tracer atom in the presence of a solute–vacancy interaction of arbitrary range. Studying model cases in order to enlighten basic trends, the conclusions can be stated as follows:

* taking into account the full range of the solute-vacancy interaction may bring noticeable changes to the value of the solute correlation factor;

* the looser the structure, the larger the change which will be directly reflected in the value of solute diffusivity;

* the solute correlation factor depends very tightly on the pattern of the jump barriers which is chosen to describe the vacancy jumps in the vicinity of the solute atom;

* taking into account the interaction up to the 3rd neighbour shell is mandatory to obtain the solute diffusivity in BCC and FCC structures with a good precision.

If a thorough ab-initio evaluation of all these barriers is highly desirable, it is unfortunately rarely available in former publications. It was shown that approximations often used to overcome this lack of information must be chosen with care. A model case was even found in which the correlation factor turns out to be independent of the interaction range.

The examination of dilute systems recently studied shows that, in agreement with the general models, the interactions in the first three neighbour shells dictate the final value with a good precision. The main improvement of the modelling comes from dropping the restrictive assumption which imposes an equal value to the jump frequencies which dissociate the vacancy-solute pair in 1st neighbour position.

## Bibliography

[1] Chandrasekhar S., Reviews of Modern Physics **15** (1943), p. 1.

[2] Bardeen J . and Herring C., Atom movements, J.H. Hollomon Ed. (American Society for Metals, Cleveland,1951), p. 87; Imperfections in nearly perfect crystals, W. Shockley Ed. (Wiley, NY) 1952, p. 261-288.

[3] Le Claire A.D. in 'Physical Chemistry: an advanced treatise', H. Eyring, D. Henderson, W.Jost Ed. (New York 1970), Vol. X, chap. 5, p. 261.

[4] Mullen J.G., Physical Review **124** (1961), p. 1723.

[5] Manning J.R., Physical Review **128** (1962), p. 2169; Physical Review **136** (1964), p. A1758.

[6] Weiss G.H., Rubin R.J., Advances in Chemical Physics 52 (1983), p. 363.

[7] Schoën A.H. and Lowen R.W., Bulletin of the American Physical Society **5** (1960), p. 280.

[8] Compaan K., Haven Y., Transactions of the Faraday Society **52** (1956), p. 786.
Compaan K., Haven Y., Transactions of the Faraday Society **54** (1958), p. 1498.

[9] Montet G., Physical Review **B7** (1973), p. 650.

[10] Besnoit P., Bocquet J.L., Lafore P., Acta Metallurgica **25** (1977), p. 265.

[11] Ishioka S., Koiwa M., Philosophical Magazine **A37** (1978), p. 517.
Koiwa M., Ishioka S., Philosophical Magazine **A47** (1983), p. 927.

[12] Sholl C.A., J. Physics C : Solide State Phys. **14** (1981), p. 2723.
Sholl C.A., Philosophical Magazine **A65** (1992), p. 749.

[13] Howard R.E., Manning J.R., Physical Review **154** (1967), p. 561.

[14] Le Claire A.D., Philosophical Magazine **21** (1970), p. 819.
Jones M.J., Le Claire A.D., Philosophical Magazine **26** (1972), p. 1191.

[15] Doan N.V., Journal of Physics and Chemistry of Solids **33** (1972), p. 2161.
Doan N.V., Bocquet J.L., Thin Solid Films **25** (1975), p. 15.

[16] Heumann Th., Journal of Physics F: Metals Physics **9** (1979), p. 1997.
Hoshino K., Iijima Y., Hirano K., Philosophical Magazine **A44** (1979), p. 961.

[17] Anthony T.R., Proc. Conf. 'Atomic Transport in Solids and Liquids' held at Marstrand 1970, Lodding A. , Lagerwall T. Ed. (Tübingen Zeitschrift für Naturforschung,1971), p. 138.
Anthony T.R., in 'Diffusion in Solids: Recent Developments', Nowick A.S., Burton J.J. Ed. (New York Academic Press 1975), p. 353.

[18] Beaman D.R., Balluffi R.W., Simmons R.O., Physical Review **134** (1964), p. A532.
Beaman D.R., Balluffi R.W., Physical Review **137** (1965), p. A917.

[19] Bocquet J.L., Brébec G., Limoge Y., in 'Physical Metallurgy' 4th revised and enlarged edition, R.W. Cahn and P. Haase, Ed. (North Holland 1996) , chap. 7, p. 536.

[20] Agarwala R.P., Materials Science Forum **1** (1984), p. 15.

[21] Tucker J.D., Najafabadi R., Allen T.R., Morgan D., Journal of Nuclear Materials **405** (2010), p.216.

[22] Soisson F., Fu C.C., Physical Review **B76** (2007), 214102.

[23] Bocquet J.L., Philosophical Magazine **A47** (1983), p. 547 ; Philosophical Magazine **A63** (1991), p. 157.

[24] Koiwa M., Ishioka S., Philosophical Magazine **A47** (1983), p. 927.

[25] Howard R.E., Physical Review **144** (1966), p. 650.

[26] Kang H.C., Weinberg W.H., Journal of Chemical Physics **90** (1989), p.2824.

[27] Garnier T., Manga V.R., Trinkle D.R., Bellon P., Physical Review **B88** (2013) 134108.

[28] Garnier T., Nastar M., Bellon P., Trinkle D.R., Physical Review **B88** (2013) 134201.
[29] Bocquet J.L., Defect and Diffusion Forum **203-205** (2002), p. 81.

## Appendix A: Special handling of $LL_O$

We start from Eq. (13) established in the main section:

$$LL_l - \sum_{\{R_j\}} \left[ \sum_{\{\omega_{1V}\}} \delta(R_j + \omega_{1V} - \{R_k\}) I_{R_l - R_j} \left( LL_j (W_O - W_{j\to k}) + LL_k (W_{k\to j} - W_O) \right) \right] = I_{CIR_l}$$

The equation used to define $LL_O$ for $p \neq 0$ is obtained for $l = 0$ :

$$LL_O - \sum_{\{R_j\}} I_{-R_j} \left[ \sum_{\{\omega_{1V}\}} \delta(R_j + \omega_{1V} - \{R_k\}) \left( LL_j (W_O - W_{j\to k}) + LL_k (W_{k\to j} - W_O) \right) \right]$$

$$= \frac{1}{V_{ZB}} \int_{V_{ZB}} \frac{CI(k)}{(p + Deno)} d_3k$$

In the left hand side, we extract the only contributions containing $LL_0$.
The first one is provided by $R_j = 0$. In the square bracket a contribution is brought whenever $\omega_{1V}$ points towards a specific site; it happens z times, since the z first neighbour of the origin are specific sites, thus giving rise to z equivalent terms:

$$-I_O \left[ \sum_{\{\omega_{1V}\}} \delta(\omega_{1V} - \{R_k\}) \left( LL_O W_O + LL_k (W_{k\to 0} - W_O) \right) \right] \to -zLL_O W_O I_O$$

The second contribution is provided by the specific sites which are first neighbours of the origin $\{R_j\} = \{\omega_{1V}\}$ : the summation in the square bracket is then restrained to one vector $\omega'_{1V}$ such that $\omega_{1V} + \omega'_{1V} = 0$ :

$$-\sum_{\{\omega_{1V}\}} I_{-\omega_{1V}} \left[ \sum_{\{\omega'_{1V}\}} \delta(\omega_{1V} + \omega'_{1V}) \left( LL_{\omega_{1V}} (W_O - W_{j\to 0}) - LL_0 W_O \right) \right]$$

$$\to +LL_O W_O \sum_{\{\omega_{1V}\}} I_{-\omega_{1V}} = LL_O \frac{1}{V_{ZB}} \int_{V_{ZB}} \frac{zW_O - Deno}{(p + Deno)} d_3k \to +zLL_O W_O I_O - LL_O [1 - pI_O]$$

The only term remaining after summation is $pI_O LL_O$ in which the 'p' factor cancels out the divergence of $LL_O$ at $p = 0$.

In all other equations used to define other probabilities for $l \neq 0$, the contributions are provided by the same values of the subscripts.
The first term provided by $R_j = 0$ is :

$$-\sum_{\{\omega_{1V}\}} \delta(\omega_{1V} - \{R_k\}) I_{R_l} \left( LL_O W_O + LL_k (W_{k\to 0} - W_O) \right) \to -zLL_O W_O I_{R_l}$$

The second contribution from $\{R_j\} = \{\omega_{1V}\}$ becomes :

$$+LL_O W_O \sum_{\{\omega_{1V}\}} \sum_{\{\omega'_{1V}\}} \delta(\omega_{1V} + \omega'_{1V}) I_{R_l - \omega_{1V}} = +LL_O W_O \sum_{\{\omega_{1V}\}} I_{R_l - \omega_{1V}} = +LL_O W_O \frac{1}{V_{ZB}} \int_{V_{ZB}} \frac{e^{ikR_l} \sum_{\{\omega_{1V}\}} e^{-ik\omega_{1V}}}{(p + Deno)} d_3k$$

The summation in the numerator of the integral is nothing but the denominator apart from a constant term:

$$= LL_O \frac{1}{V_{ZB}} \int_{V_{ZB}} \frac{e^{ikR_l}(zW_O - Deno)}{(p + Deno)} d_3k = zW_O LL_O I_{R_l} - LL_O \frac{1}{V_{ZB}} \int_{V_{ZB}} \frac{e^{ikR_l}(p + Deno) - p}{(p + Deno)} d_3k$$

$$= zW_O LL_O I_{R_l} - LL_O \frac{1}{V_{ZB}} \int_{V_{ZB}} e^{ikR_l} d_3k + pLL_O I_{R_l}$$

The integral is zero and the summation is reduced to $zW_O LL_O I_{R_l} + pLL_O I_{R_l}$ .

The summation of all the contributions yields $pLL_O I_{R_l}$ .
As a result, the divergent quantity $LL_O$ which behaves like $1/p$ when $p \to 0$ always appears in the equations with a multiplying factor proportional to $p$ .

## Appendix B: Establishing the corrective terms in the transport equation for the five frequency model for FCC lattice

Starting from a given site of $S_4$ belonging to $\{R_4^+\}$, the subsets which can be reached in one jump are $\{R_0\}$, $\{R_1\}$, $\{R_2\}$, $\{R_4^+\}$, $\{R_5^+\}$, $\{R_6^+\}$, $\{R_7^+\}$ and $\{R_8^+\}$ with the corresponding values of $nlien_{i\to j}$ displayed below in Table B1

| $nlien_{i\to j}$ $j\rightarrow$ $i\downarrow$ | 0 | 1 | 2 | 3 | 4 | 5 | 6 | 7 | 8 |
|---|---|---|---|---|---|---|---|---|---|
| 4 | 1 | 2 | 1 | 0 | 2 | 1 | 2 | 2 | 1 |
| 5 | 0 | 0 | 0 | 0 | 4 | 0 | 0 | 4 | 0 |
| 6 | 0 | 1 | 1 | 1 | 1 | 0 | 1 | 1 | 1 |
| 7 | 0 | 0 | 0 | 0 | 2 | 1 | 2 | 0 | 2 |
| 8 | 0 | 0 | 0 | 0 | 1 | 0 | 2 | 2 | 0 |

*Table B1. Multiplicity of links between subsets in the five frequency model.*

**Corrective term for $S_4$ :**

$W_O$ must then be replaced by the corresponding outward jump frequencies : $W_{4\to0}$, $W_{4\to1}$, $W_{4\to2}$, $W_{4\to4}$, $W_{4\to5}$, $W_{4\to6}$, $W_{4\to7}$, $W_{4\to8}$ respectively, while taking due account of the multiplicity. Hence the first part of the correction:

$$+LL_4^+(12W_O - W_{4\to0} - 2W_{4\to1} - W_{4\to2} - 2W_{4\to4} - W_{4\to5} - 2W_{4\to6} - 2W_{4\to7} - W_{4\to8}).$$

The second part of the correction comes from the inward frequencies from the subsets associated to a non-zero occupancy, namely:

$$+2LL_4^+(W_{4\to4} - W_O) + LL_5^+(W_{5\to4} - W_O) + 2LL_6^+(W_{6\to4} - W_O) + 2LL_7^+(W_{7\to4} - W_O) + LL_8^+(W_{8\to4} - W_O)$$

The corrective terms being identical for all sites belonging to $\{R_4^+\}$, the Fourier transform of the transport equation brings a contribution:

$$+f_4^+\left\{\begin{array}{l} LL_4^+(12W_O - W_{4\to0} - 2W_{4\to1} - W_{4\to2} - 2W_{4\to4} - W_{4\to5} - 2W_{4\to6} - 2W_{4\to7} - W_{4\to8}) \\ +2LL_4^+(W_{4\to4} - W_O) + LL_5^+(W_{5\to4} - W_O) + 2LL_6^+(W_{6\to4} - W_O) \\ +2LL_7^+(W_{7\to4} - W_O) + LL_8^+(W_{8\to4} - W_O) \end{array}\right\}$$

The same work has to be made on the negative side of x-axis for the sites belonging to $\{R_4^-\}$. The definition of the subsets preserving their neighbourhood relationships, it can be immediately written down as:

$$f_4^-\left\{\begin{array}{l} LL_4^-(12W_O - W_{4\to0} - 2W_{4\to1} - W_{4\to2} - 2W_{4\to4} - W_{4\to5} - 2W_{4\to6} - 2W_{4\to7} - W_{4\to8}) \\ +2LL_4^-(W_{4\to4} - W_O) + LL_5^-(W_{5\to4} - W_O) + 2LL_6^-(W_{6\to4} - W_O) \\ +2LL_7^-(W_{7\to4} - W_O) + LL_8^-(W_{8\to4} - W_O) \end{array}\right\}$$

Remembering that $LL_i^- = -LL_i^+$, the final expression of the correction comes out as:

$$f_4\begin{Bmatrix} LL_4^+(12W_O - W_{4\to0} - 2W_{4\to1} - W_{4\to2} - 2W_{4\to4} - W_{4\to5} - 2W_{4\to6} - 2W_{4\to7} - W_{4\to8}) \\ +2LL_4^+(W_{4\to4} - W_O) + LL_5^+(W_{5\to4} - W_O) + 2LL_6^+(W_{6\to4} - W_O) \\ +2LL_7^+(W_{7\to4} - W_O) + LL_8^+(W_{8\to4} - W_O) \end{Bmatrix}$$

To comply with the standard notations of the five frequency model, the correction is transformed with the help of Table 2 of the main section and reads:

$$+f_4\begin{Bmatrix} LL_4^+(10W_O - W_2 - 2W_1 - 7W_3) + LL_5^+(W_4 - W_O) + 2LL_6^+(W_4 - W_O) \\ +2LL_7^+(W_4 - W_O) + LL_8^+(W_4 - W_O) \end{Bmatrix} \quad \text{(B1)}$$

In the same way, the formal expressions of other corrective terms are given below, together with the formulations using the notations of the standard model.

**Corrective term for $S_5$ :**

$$+f_5\begin{Bmatrix} LL_5^+(8W_O - 4W_{5\to4} - 4W_{5\to7}) \\ +4LL_4^+(W_{4\to5} - W_O) + 4LL_7^+(W_{7\to5} - W_O) \end{Bmatrix}$$

or $+f_5\left\{4LL_5^+(W_O - W_4) + 4LL_4^+(W_3 - W_O)\right\}$ (B2)

**Corrective term for $S_6$ :**

$$+f_6\begin{Bmatrix} LL_6^+(7W_O - W_{6\to1} - W_{6\to2} - W_{6\to3} - W_{6\to4} - W_{6\to6} - W_{6\to7} - W_{6\to8}) \\ +LL_4^+(W_{4\to6} - W_O) + LL_6^+(W_{6\to6} - W_O) + LL_7^+(W_{7\to6} - W_O) + LL_8^+(W_{8\to6} - W_O) \end{Bmatrix}$$

or $f_6\left\{2LL_6^+(W_O - W_4) + LL_4^+(W_3 - W_O)\right\}$ (B3)

**Corrective term for $S_7$ :**

$$+f_7\begin{Bmatrix} LL_7^+(7W_O - 2W_{7\to4} - W_{7\to5} - 2W_{7\to6} - 2W_{7\to8}) \\ +2LL_4^+(W_{4\to7} - W_O) + LL_5^+(W_{5\to7} - W_O) + 2LL_6^+(W_{6\to7} - W_O) + 2LL_8^+(W_{8\to7} - W_O) \end{Bmatrix}$$

or $+f_7\left\{2LL_7^+(W_O - W_4) + 2LL_4^+(W_3 - W_O)\right\}$ (B4)

**Corrective term for $S_8$ :**

$$+f_8\begin{Bmatrix} LL_8^+(5W_O - W_{8\to4} - 2W_{8\to6} - 2W_{8\to7}) \\ +LL_4^+(W_{4\to8} - W_O) + 2LL_6^+(W_{6\to8} - W_O) + 2LL_7^+(W_{7\to8} - W_O) \end{Bmatrix}$$

or $+f_8\left\{LL_8^+(W_O - W_4) + LL_4^+(W_3 - W_O)\right\}$ (B5)

Since we can safely set $p$ equal to zero before performing the inverse Fourier transform, the final formulation of the transport equation is recast in order to yield the coefficients $g_i(k)/D_0$ of Eq. (28) with $W_i^{'} = \dfrac{W_i - W_O}{W_O}$:

$$FLL(k,p)\Big|_{p=0} = \frac{ci(k)}{W_O D_0} + LL_4^+ \frac{\left[-(W_2^{'}+2W_1^{'}+7W_3^{'})f_4 + W_3^{'}(4f_5+f_6+2f_7+f_8)\right]}{D_0}$$
$$+ LL_5^+ \frac{\left[-4W_4^{'}f_5 + W_4^{'}f_4\right]}{D_0} + LL_6^+ \frac{\left[-2W_4^{'}f_6 + 2W_4^{'}f_4\right]}{D_0} \quad \text{(B6)}$$
$$+ LL_7^+ \frac{\left[-2W_4^{'}f_7 + 2W_4^{'}f_4\right]}{D_0} + LL_8^+ \frac{\left[-W_4^{'}f_8 + W_4^{'}f_4\right]}{D_0}$$

The expressions of the functions $g_j(k)$ introduced in Eqs. 24 are then given by:

$$g_4(k) = -(W_2^{'}+2W_1^{'}+7W_3^{'})f_4 + W_3^{'}\left[4f_5+f_6+2f_7+f_8\right]$$
$$g_5(k) = -4W_4^{'}f_5 + W_4^{'}f_4$$
$$g_6(k) = -2W_4^{'}f_6 + 2W_4^{'}f_4$$
$$g_7(k) = -2W_4^{'}f_7 + 2W_4^{'}f_4$$
$$g_8(k) = -W_4^{'}f_8 + W_4^{'}f_4$$

which yields the final expression of the doubly transformed function in Eqs. 24.

*Establishing the first equation:*

$$LL_4^+ = -\int \frac{f_4}{8} FLL(k,p)\Big|_{p=0} d_3k \Rightarrow$$
$$LL_4^+\left\{1+(W_2^{'}+2W_1^{'}+7W_3^{'})M_{441} - W_3^{'}M_{442}\right\} - LL_5^+W_4^{'}M_{45} - LL_6^+W_4^{'}M_{46} \quad \text{(B7)}$$
$$- LL_7^+W_4^{'}M_{47} - LL_8^+W_4^{'}M_{48} = M_4 / W_O$$

with the following definitions of the coefficients in which we kept only the real parts $F_i$ and $CI$ of the imaginary functions $f_i$ and of the initial condition $ci(k)$:

$$M_{441} = \frac{1}{V_{ZB}}\int \frac{F_4F_4}{8D_0} d_3k \qquad M_{442} = \frac{1}{V_{ZB}}\int \frac{F_4\left[4F_5+F_6+2F_7+F_8\right]}{8D_0} d_3k$$
$$M_{45} = \frac{1}{V_{ZB}}\int \frac{F_4}{8}\frac{\left[-4F_5+F_4\right]}{D_0} d_3k \qquad M_{46} = \frac{1}{V_{ZB}}\int \frac{F_4}{8}\frac{\left[-2F_6+2F_4\right]}{D_0} d_3k$$
$$M_{47} = \frac{1}{V_{ZB}}\int \frac{F_4}{8}\frac{\left[-2F_7+2F_4\right]}{D_0} d_3k \qquad M_{48} = \frac{1}{V_{ZB}}\int \frac{F_4}{8}\frac{\left[-F_8+F_4\right]}{D_0} d_3k$$
$$M_4 = \frac{1}{V_{ZB}}\int \frac{F_4}{8}\frac{CI(k)}{D_0} d_3k$$

*Establishing the second equation:*

$$LL_5^+ = -\int \frac{f_5}{2} FLL(k,p) \Rightarrow$$

$$-LL_4^+\left\{-(W_2^{'}+2W_1^{'}+7W_3^{'})M_{541}+W_3^{'}M_{542}\right\}+LL_5^+\left\{1-W_4^{'}M_{55}\right\}-LL_6^+W_4^{'}M_{56} -LL_7^+W_4^{'}M_{57}-LL_8^+W_4^{'}M_{58}=M_5/W_O \qquad \text{(B8)}$$

with :

$$M_{541}=\frac{1}{V_{ZB}}\int \frac{F_5}{2}\frac{F_4}{D_0}d_3k \qquad M_{542}=\frac{1}{V_{ZB}}\int \frac{F_5}{2}\frac{\left[4F_5+F_6+2F_7+F_8\right]}{D_0}d_3k$$

$$M_{55}=\frac{1}{V_{ZB}}\int \frac{F_5}{2}\frac{\left[-4F_5+F_4\right]}{D_0}d_3k \qquad M_{56}=\frac{1}{V_{ZB}}\int \frac{F_5}{2}\frac{\left[-2F_6+2F_4\right]}{D_0}d_3k$$

$$M_{57}=\frac{1}{V_{ZB}}\int \frac{F_5}{2}\frac{\left[-2F_7+2F_4\right]}{D_0}d_3k \qquad M_{58}=\frac{1}{V_{ZB}}\int \frac{F_5}{2}\frac{\left[-F_8+F_4\right]}{D_0}d_3k$$

$$M_5=\frac{1}{V_{ZB}}\int \frac{F_5}{2}\frac{CI(k)}{D_0}d_3k$$

*Establishing the third equation:*

$$LL_6^+ = -\int \frac{f_6}{16} FLL(k,p) \Rightarrow$$

$$-LL_4^+\left\{-(W_2^{'}+2W_1^{'}+7W_3^{'})M_{641}+W_3^{'}M_{642}\right\}-LL_5^+W_4^{'}M_{65}+LL_6^+\left\{1-W_4^{'}M_{66}\right\} -LL_7^+W_4^{'}M_{67}-LL_8^+W_4^{'}M_{68}=M_6/W_O \qquad \text{(B9)}$$

with :

$$M_{641}=\frac{1}{V_{ZB}}\int \frac{F_6}{16}\frac{F_4}{D_0}d_3k \qquad M_{642}=\frac{1}{V_{ZB}}\int \frac{F_6}{16}\frac{\left[4F_5+F_6+2F_7+F_8\right]}{D_0}d_3k$$

$$M_{65}=\frac{1}{V_{ZB}}\int \frac{F_6}{16}\frac{\left[-4F_5+F_4\right]}{D_0}d_3k \qquad M_{66}=\frac{1}{V_{ZB}}\int \frac{F_6}{16}\frac{\left[-2F_6+2F_4\right]}{D_0}d_3k$$

$$M_{67}=\frac{1}{V_{ZB}}\int \frac{F_6}{16}\frac{\left[-2F_7+2F_4\right]}{D_0}d_3k \qquad M_{68}=\frac{1}{V_{ZB}}\int \frac{F_6}{16}\frac{\left[-F_8+F_4\right]}{D_0}d_3k$$

$$M_6=\frac{1}{V_{ZB}}\int \frac{F_6}{16}\frac{CI(k)}{W_0D_0}d_3k$$

*Establishing the fourth equation:*

$$LL_7^+ = -\int \frac{f_7}{8} FLL(k,p) \Rightarrow$$

$$-LL_4^+\left\{-(W_2^{'}+2W_1^{'}+7W_3^{'})M_{741}+W_3^{'}M_{742}\right\}-LL_5^+W_4^{'}M_{75}-LL_6^+W_4^{'}M_{76} +LL_7^+\left\{1-W_4^{'}M_{77}\right\}-LL_8^+W_4^{'}M_{78}=M_7/W_O \qquad \text{(B10)}$$

with :

$$M_{741}=\frac{1}{V_{ZB}}\int \frac{F_7}{8}\frac{F_4}{D_0}d_3k \qquad M_{742}=\frac{1}{V_{ZB}}\int \frac{F_7}{8}\frac{[4F_5+F_6+2F_7+F_8]}{D_0}d_3k$$

$$M_{75}=\frac{1}{V_{ZB}}\int \frac{F_7}{8}\frac{[-4F_5+F_4]}{D_0}d_3k \qquad M_{76}=\frac{1}{V_{ZB}}\int \frac{F_7}{8}\frac{[-2F_6+2F_4]}{D_0}d_3k$$

$$M_{77}=\frac{1}{V_{ZB}}\int \frac{F_7}{8}\frac{[-2F_7+2F_4]}{D_0}d_3k \qquad M_{78}=\frac{1}{V_{ZB}}\int \frac{F_7}{8}\frac{[-F_8+F_4]}{D_0}d_3k$$

$$M_7=\int \frac{F_7}{8}\frac{CI(k)}{D_0}d_3k$$

*Establishing the fifth equation:*

$$LL_8^+=-\int \frac{f_8}{8}FLL(k,p)\Rightarrow$$

$$-LL_4^+\left\{-(W_2^{'}+2W_1^{'}+7W_3^{'})M_{841}+W_3^{'}M_{842}\right\}-LL_5^+W_4^{'}M_{85}-LL_6^+W_4^{'}M_{86}$$
$$-LL_7^+W_4^{'}M_{87}+LL_8^+\left\{1-W_4^{'}M_{88}\right\}=M_8/W_O \qquad \text{(B11)}$$

with:

$$M_{841}=\frac{1}{V_{ZB}}\int \frac{F_8}{8}\frac{F_4}{D_0}d_3k \qquad M_{842}=\frac{1}{V_{ZB}}\int \frac{F_8}{8}\frac{[4F_5+F_6+2F_7+F_8]}{D_0}d_3k$$

$$M_{85}=\frac{1}{V_{ZB}}\int \frac{F_8}{8}\frac{[-4F_5+F_4]}{D_0}d_3k \qquad M_{86}=\frac{1}{V_{ZB}}\int \frac{F_8}{8}\frac{[-2F_6+2F_4]}{D_0}d_3k$$

$$M_{87}=\frac{1}{V_{ZB}}\int \frac{F_8}{8}\frac{[-2F_7+2F_4]}{D_0}d_3k \qquad M_{88}=\frac{1}{V_{ZB}}\int \frac{F_8}{8}\frac{[-F_8+F_4]}{D_0}d_3k$$

$$M_8=\frac{1}{V_{ZB}}\int \frac{F_8}{8}\frac{CI(k)}{D_0}d_3k$$

Indeed, the integrals introduced above are not independent.
Obviously: $M_{441}=4M_4$ because $CI(k)=F_4/4$ and $-f_if_j=F_iF_j$
In addition, there is a linear relationship between $M_{441}$ and $M_{442}$:

$$M_{441}=\frac{1}{V_{ZB}}\int_{V_{ZB}}\frac{F_4F_4}{8D_0}d_3k=\frac{1}{V_{ZB}}\int_{V_{ZB}}\frac{16s_x^2(c_y+c_z)^2}{8D_0}d_3k=\frac{1}{V_{ZB}}\int_{V_{ZB}}\frac{2s_x^2(c_y+c_z)^2}{D_0}d_3k$$

and

$$M_{442}=\frac{1}{V_{ZB}}\int_{V_{ZB}}\frac{[4F_5+F_6+2F_7+F_8]F_4}{8D_0}d_3k$$
$$=\frac{1}{V_{ZB}}\int_{V_{ZB}}\frac{16s_x^2(c_y+c_z)^2[-D_0+10]}{8D_0}d_3k$$
$$=-\frac{1}{V_{ZB}}\int_{V_{ZB}}2s_x^2(c_y+c_z)^2d_3k+\frac{1}{V_{ZB}}\int_{V_{ZB}}\frac{20s_x^2(c_y+c_z)^2}{D_0}d_3k$$
$$=-1+10M_{441}$$

For the similar reasons: $M_{541} = 4M_5 \qquad M_{641} = 4M_6 \qquad M_{741} = 4M_7 \qquad M_{841} = 4M_8$
In addition:

$$M_{541} = \int \frac{F_5}{2}\frac{F_4}{D_0} d_3k = \int \frac{2s_{2x}}{2}\frac{4s_x(c_y + c_z)}{D_0} d_3k = \int \frac{8s_x^2 c_x(c_y + c_z)}{D_0} d_3k$$

$$M_{542} = \int \frac{F_5}{2}\frac{\left[4F_5 + F_6 + 2F_7 + F_8\right]}{D_0} d_3k = \int \frac{2s_{2x}}{2}\frac{4s_x(c_y + c_z)\left[-D_0 + 10\right]}{D_0} d_3k$$

$$= \int \frac{8s_x^2 c_x(c_y + c_z)\left[-D_0 + 10\right]}{D_0} d_3k = -\int 8s_x^2 c_x(c_y + c_z) d_3k + \int \frac{80s_x^2 c_x(c_y + c_z)}{D_0} d_3k$$

$$= 10M_{541}$$

and further : $M_{642} = 10M_{641} \qquad M_{742} = 10M_{741} \qquad M_{842} = 10M_{841}$

The matrix of coefficients for the linear system, together with the column of right hand sides, is displayed in Table B2 below.

| | | | | | |
|---|---|---|---|---|---|
| $\left\{(W_2^{'} + 2W_1^{'} + 7W_3^{'})M_{441} - W_3^{'}M_{442}\right\} + 1$ | $-W_4^{'}M_{45}$ | $-W_4^{'}M_{46}$ | $-W_4^{'}M_{47}$ | $-W_4^{'}M_{48}$ | $M_4 / W_O$ |
| $\left\{(W_2^{'} + 2W_1^{'} + 7W_3^{'})M_{541} - W_3^{'}M_{542}\right\}$ | $1 - W_4^{'}M_{55}$ | $-W_4^{'}M_{56}$ | $-W_4^{'}M_{57}$ | $-W_4^{'}M_{58}$ | $M_5 / W_O$ |
| $\left\{(W_2^{'} + 2W_1^{'} + 7W_3^{'})M_{641} - W_3^{'}M_{642}\right\}$ | $-W_4^{'}M_{65}$ | $1 - W_4^{'}M_{66}$ | $-W_4^{'}M_{67}$ | $-W_4^{'}M_{68}$ | $M_6 / W_O$ |
| $\left\{(W_2^{'} + 2W_1^{'} + 7W_3^{'})M_{741} - W_3^{'}M_{742}\right\}$ | $-W_4^{'}M_{75}$ | $-W_4^{'}M_{76}$ | $1 - W_4^{'}M_{77}$ | $-W_4^{'}M_{78}$ | $M_7 / W_O$ |
| $\left\{(W_2^{'} + 2W_1^{'} + 7W_3^{'})M_{841} - W_3^{'}M_{842}\right\}$ | $-W_4^{'}M_{85}$ | $-W_4^{'}M_{86}$ | $-W_4^{'}M_{87}$ | $1 - W_4^{'}M_{88}$ | $M_8 / W_O$ |

*Table B2 : Matrix of coefficients for linear system*

The determinant is expressed with the cofactors $D_{i4}$ $(i = 4, 8)$ of the first column:

$$\Delta = \left\{1 + (W_2^{'} + 2W_1^{'} + 7W_3^{'})M_{441} - W_3^{'}M_{442}\right\} D_{44} + \left\{(W_2^{'} + 2W_1^{'} + 7W_3^{'})M_{541} - W_3^{'}M_{542}\right\} D_{54}$$
$$+ \left\{(W_2^{'} + 2W_1^{'} + 7W_3^{'})M_{641} - W_3^{'}M_{642}\right\} D_{64} + \left\{(W_2^{'} + 2W_1^{'} + 7W_3^{'})M_{741} - W_3^{'}M_{742}\right\} D_{74}$$
$$+ \left\{(W_2^{'} + 2W_1^{'} + 7W_3^{'})M_{841} - W_3^{'}M_{842}\right\} D_{84}$$
$$= D_{44} + (W_2^{'} + 2W_1^{'} + 7W_3^{'})(M_{441}D_{44} + M_{541}D_{54} + M_{641}D_{64} + M_{741}D_{74} + M_{841}D_{84})$$
$$- W_3^{'}(M_{442}D_{44} + M_{542}D_{54} + M_{642}D_{64} + M_{742}D_{74} + M_{842}D_{84})$$

Obviously, cofactors are functions of $W_4^{'} = (W_4 - W_O)/W_O$ only.

The average product $< x_1x_2 >$ is given by $< x_1x_2 > = -4W_2LL_4^+ = -4W_2Num_4^+/\Delta$ with

$$Num_4^+ = (M_4D_{44} + M_5D_{54} + M_6D_{64} + M_7D_{74} + M_8D_{84})/W_O$$
$$= (M_{441}D_{44} + M_{451}D_{54} + M_{461}D_{64} + M_{471}D_{74} + M_{481}D_{84})/(4W_O)$$

and the final expression of the correlation factor is given by:

$$f = \frac{1 + < x_1x_2 >}{1 - < x_1x_2 >} = \frac{\Delta - 4W_2Num_4^+}{\Delta + 4W_2Num_4^+}$$

To recast the expression under the standard form displayed in the previous publications using un-primed frequencies, it is worth noticing the following.

* frequency $W_2$ does vanish from the numerator:

$$\Delta - 4W_2Num_4^+ = \Delta - 4W_O(1 + W_2^{'})Num_4^+$$
$$= D_{44} + (W_2^{'} + 2W_1^{'} + 7W_3^{'})(M_{441}D_{44} + M_{541}D_{54} + M_{641}D_{64} + M_{741}D_{74} + M_{841}D_{84})$$
$$- W_3^{'}(M_{442}D_{44} + M_{542}D_{54} + M_{642}D_{64} + M_{742}D_{74} + M_{842}D_{84})$$
$$- (1 + W_2^{'})(M_{441}D_{44} + M_{541}D_{54} + M_{641}D_{64} + M_{741}D_{74} + M_{841}D_{84})$$
$$= (2W_1^{'} + 7W_3^{'})(M_{441}D_{44} + M_{541}D_{54} + M_{641}D_{64} + M_{741}D_{74} + M_{841}D_{84})$$
$$- W_3^{'}(M_{442}D_{44} + M_{542}D_{54} + M_{642}D_{64} + M_{742}D_{74} + M_{842}D_{84})$$
$$+ D_{44} - (M_{441}D_{44} + M_{541}D_{54} + M_{641}D_{64} + M_{741}D_{74} + M_{841}D_{84})$$

* turning back to un-primed frequencies, the numerator of the correlation factor becomes:

$$Num_4^+ = 2\frac{W_1}{W_O}(M_{441}D_{44} + M_{541}D_{54} + M_{641}D_{64} + M_{741}D_{74} + M_{841}D_{84})$$
$$+ 7\frac{W_3}{W_O}\begin{Bmatrix}(M_{441}D_{44} + M_{541}D_{54} + M_{641}D_{64} + M_{741}D_{74} + M_{841}D_{84}) \\ -(M_{442}D_{44} + M_{542}D_{54} + M_{642}D_{64} + M_{742}D_{74} + M_{842}D_{84})/7\end{Bmatrix}$$
$$+ (1 + M_{442} - 10M_{441})D_{44} + (M_{542} - 10M_{541})D_{54} + (M_{642} - 10M_{641})D_{64}$$
$$+ (M_{742} - 10M_{741})D_{74} + (M_{842} - 10M_{841})D_{84}$$

in which the terms independent of frequencies cancel out thanks to the relationships which were established above.

At last the correlation factor can be recast into the standard form used for the five frequency model :

$$f_{B*} = \frac{2W_1 + 7FW_3}{2W_1 + 2W_2 + 7FW_3}$$

where the function $7F$ depends only on $W_4^{'}$ , i.e. on the ratio $W_4 / W_O$

$$7F\left(W_4^{'}\right) = 7 - \frac{(10M_{441} - 1)D_{44} + 10M_{541}D_{54} + 10M_{641}D_{64} + 10M_{741}D_{74} + 10M_{841}D_{84}}{(M_{441}D_{44} + M_{541}D_{54} + M_{641}D_{64} + M_{741}D_{74} + M_{841}D_{84})}$$

$$= -3 + \frac{D_{44}}{M_{441}D_{44} + M_{541}D_{54} + M_{641}D_{64} + M_{741}D_{74} + M_{841}D_{84}}$$

One remarkable result is the value which corresponds to $W_4^{'} = 0$ or $W_4 / W_0 = 1$. In this case, all the cofactors are equal to zero but $D_{44}$ which is equal to unity. Hence:

$7F\left(W_4 / W_0 = 1\right) = -3 + \frac{1}{4M_4} = 5.1512835$ in agreement with all the previous evaluations which were done in the past.[5,24]

The integrals are calculated with a superposition of Gaussian quadratures over $[-\pi, +\pi]$ on the 3 axis. The interval is divided into $n_{int}$ sub-intervals and in each of them the abscissas and weights of the Gaussian quadrature are applied to $n_{points}$. Several couples ($n_{int}$ , $n_{points}$) were used while keeping the product $n_{int}*n_{points}$ equal to 1024. The decimal digits displayed in Table B3 are the only ones to be shared by all evaluations.

Our analytical formulation of the function 7F looks different from the usual one, because the cofactors are here expressed as functions of the primed frequency $W_4^{'} = (W_4 - W_O) / W_O$, which appears as a more natural variable in the transport equation, instead of $W_4 / W_O$. But the numerical solution of the linear system gives of course a final result which is identical to the most precise evaluation published so far within 8 digits at least. [24]

| $i \downarrow$ | $M_{i41}$ | $M_{i5}$ | $M_{i6}$ | $M_{i7}$ | $M_{i8}$ |
|---|---|---|---|---|---|
| 4 | 1.226800660D-01 | 5.768642893D-02 | 1.5786196752D-01 | 1.578619675D-01 | 9.212028963D-02 |
| 5 | 6.4993637077D-02 | -3.43781570D-01 | 3.9969164488D-02 | -2.5062653657D-02 | 2.3891456749D-02 |
| 6 | 2.1874541124D-02 | -6.2998629D-04 | -1.892626651D-01 | -1.043043296D-02 | -3.680953757D-03 |
| 7 | 4.3749082246D-02 | -3.3775881659D-02 | -2.0860865927D-02 | -1.70210352D-01 | -4.3287444760D-03 |
| 8 | 3.0559776378D-02 | -1.05424039491D-02 | -4.110242677D-02 | -3.50361006885D-02 | -8.66691638D-02 |

*Table B3 : Values of lattice integrals for the five frequency model.*

## Appendix C: Fortran code

We give hereafter a few hints about the Fortran code which was used to calculate solute correlation factors:

1/ the list of coordinates for all sites lying at a given distance from origin in spherical coordination shells is established. Sub-shells are also detected during the numbering (for instance (4,1,1) and (3,3,0), (5,1,0) and (4,3,1), (5,3,0) and (4,3,3) …etc for the FCC lattice; (5,1,1) and (3,3,3), (6,0,0) and (4,4,2), (7,1,1) and (5,5,1) …etc for the BCC lattice) and later merged into a single super-shell, defined only by its number $ic$ and its squared distance from the origin $ir2_de_ic(ic)$. Several other quantities are recorded:

$nsc(ic)$: number of sub-shells of shell number $ic$;

$nvsc(isc,ic)$ : number of sites in the sub-shell $isc \in [1, nsc(ic)]$ of shell $ic$;

$nv_couche(ic)$ : total number of sites in shell $ic$;

$ijkv_couche(i,j,ic)$ : coordinates (i=1,3) of $j^{th}$ site of shell $ic$;

$num_couche(i,j,k)$ : number $ic$ of the shell containing site $(i,j,k)$.

By definition shell $ic = 0$ is the origin.

2/ the list of all neighbour shells which can be reached in one jump from a given one is established:

$ic1v(j,ic)$ : number of the $j^{th}$ shell which can be reached from $ic$ via one first neighbour jump with $j \in [1, nv1c(ic)]$;

$idelta_couche(ic)$ : maximum difference between $ic$ and $ic1v(j,ic)$.

3/ the input parameter is the range of the solute-vacancy interaction denoted by the variable $ncouche_interac$; the total number of spherical neighbour shells to be used is $ncouche = ncouche_interac + idelta_couche\,(ncouche_interac)$.

4/ all the subsets which are necessitated by the range of interaction are numbered. The first $m_{sub} - 1 = nsubset_decal$ subsets are obtained by intersecting the $ncouche$ shells of neighbours with plane x=0. The remaining ones are obtained by intersecting planes x (>0) with the neighbour shells:

$nv_subset(ksub)$ : number of sites in subset number $ksub \in [0, M_{sub}]$, where the lower bound corresponds to the origin;

$iv_subset(i=1,3,j,ksub)$ : coordinates of $j^{th}$ site in subset $ksub$;

$num_subset(i,j,k)$ : number $ksub$ of the subset containing site (i,j,k);

$idir_subset(j,ksub)$ : number of $j^{th}$ subset which can be reached from subset $ksub$ through one first neighbour jump with $j \in [1, ndir_subset(ksub)]$ (and which can be equal to $ksub$ itself);

$nlien_{j \to k}$ : number of first neighbour jumps connecting a given site of subset j to sites belonging to subset k.

5/ the expressions of the corrective terms in the transport equation are established and computed; their formal analytical expressions can be written in dedicated output files 'FCC_transequa_correction_xx' and 'BCC_transequa_correction_xx' for a visual control whenever desired, where 'xx' stands for the two-digit-number of the subset attached to the corrective term.

6/ the knowledge of the interaction energy $E_j$ between a solute atom on the origin and a vacancy in shell $j$ helps us to define a thermodynamically consistent set of frequencies w_couche($j_{shell}$,$k_{shell}$) from shell 'j' to shell 'k' in the neighbourhood of the solute atom (the ground state being the energy of the system with a solute atom on the origin and a vacancy far away). These frequencies are then used to define the jump frequencies w_subset ($j_{sub}$,$k_{sub}$) which enter into play in the transport equation and which are now defined from subset 'j' to a neighbouring subset 'k'. It is impossible to give, once for all, a universal and explicit correspondence between the jump frequency between subsets and the jump frequency between shells because it rests on the relationships $j_{sub} \leftrightarrow j_{shell}$ and $k_{sub} \leftrightarrow k_{shell}$ which can be determined only after inspection of the two lattice sites under examination.

Further this correspondence cannot even be established once for all for a given lattice symmetry because the numbering of subsets is not conserved when increasing the interaction range for two reasons due to the connectivity of the lattice and to the changing value of $nsubset_decal$.

Let us illustrate the first point on the BCC lattice, when the solute-vacancy interaction range increases from first to second neighbour shell. Each subset is represented by its representative site $(i_1,i_2,i_3)$ together with a subscript recalling the corresponding neighbour shell for convenience:

*for a first neighbour interaction the definition runs as follows:

0≡$(0,0,0)_0$  1≡ $(0,0,2)_2$  2≡ $(0,2,2)_3$  3≡ $(1,1,1)_1$  4≡ $(2,0,0)_2$  5≡ $(2,0,2)_3$  6≡ $(2,2,2)_5$ ;

*for an interaction extending up to the second neighbour shell, the definition of the subsets becomes:

0≡ $(0,0,0)_0$  1≡ $(0,0,2)_2$  2≡ $(0,2,2)_3$  3≡ $(1,1,1)_1$  4≡ $(2,0,0)_2$  5≡ $(2,0,2)_3$  6≡$(1,1,3)_4$  7≡ $(3,1,1)_4$  8≡ $(2,2,2)_5$ .

Extending the interaction from first to second neighbour intercalates two subsets belonging to the $4^{th}$ neighbour shell, the latter being out of reach before; the numbering of subset (2,2,2) shifts from 6 to 8, although the value of $nsub_decal$ was not modified. This kind of reshuffling will take place whenever a vacancy appears in the list of the accessible neighbour shells while extending progressively the range of the interaction. This difficulty is only encountered for the BCC lattice, the density of the FCC lattice being large enough to prevent it.

In addition, extending the interaction range still farther intercalates also subsets in plane x=0 and increases the value of $nsubset_decal$, which shifts the number associated to a given subset. Subset $(2,2,2)_5$ will then be numbered 10 for an interaction range extending to $4^{th}$ neighbours, 12 for an interaction range extending to $7^{th}$ neighbours, 13 for an interaction range extending to $10^{th}$ neighbours …etc.

For this reason, a first run of the program using a ghost template of the subroutine defining the functions $f_{l_1l_2l_3}(k)$ examines the neighbourhoods, defines the subsets to be taken into consideration for a given interaction range and generates the few lines of Fortran code defining the periodic functions which enter the lattice integrals entering Eq.(22). Then it stops and starts again with a new compilation and link edition including the augmented version of the subroutine, which is now able to calculate numerically the desired integrals.

7/ the multiplicative factors of $LL_i^+$ entering into play in Eq. (22) of the main section are formally calculated. Their expression can be printed in dedicated output files ‘FCC_transequa_coeff_xx’ and ‘BCC_transequa_coeff_xx’ where ‘xx’ stands for the two-digit number ‘i’ attached to the subset under consideration.

8/ all the contributions to the coefficients entering the linear system of Eqs (24) are summed. These coefficients are made of lattice integrals $M_{ijk}$ and $M_{ij}$ multiplied by linear combinations of frequencies $W'_{j\to k}$ which are defined as transition rates between subsets. They cannot be written a priori in another way than a formal one which retains only a generic name $W'_{j\to k}$, irrespective of its numerical value (possibly zero) which depends on the detailed description of the solute-vacancy interaction under study. They can be printed in dedicated output files ‘FCC_syslin_coeff_xx_yy’ and ‘BCC_syslin_coeff_xx_yy’ for a visual check whenever desired, where ‘xx’ and ‘yy’ stand for the line and column two-digit indices or the subsets under examination.

As an illustration on the FCC lattice with an interaction ranging up to the 1st neighbours, the contributions to the corrective term appearing in the transport equation for subset $(1,1,0)_1$ corresponding to $k_{sub} = 4$ is printed in the output file named ‘FCC_transequa_correction_04_04’ with two alternative formulations using w_subset(:,:) or w_couche(:,:) as:

```
 -n_lien( 4, 8) f( 4) LLS( 4) w_subset( 4, 8)           +n_lien( 4, 8) f( 4) LLS( 8) w_subset(
8, 4)           avec n_lien( 4, 8) =  1
         equal to :
 -n_lien( 4, 8) f( 4) LLS( 4) w_couche( 1, 4)           +n_lien( 4, 8) f( 4) LLS( 8) w_couche(
4, 1)

 -n_lien( 4, 7) f( 4) LLS( 4) w_subset( 4, 7)           +n_lien( 4, 7) f( 4) LLS( 7) w_subset(
7, 4)           avec n_lien( 4, 7) =  2
         equal to :
 -n_lien( 4, 7) f( 4) LLS( 4) w_couche( 1, 3)           +n_lien( 4, 7) f( 4) LLS( 7) w_couche(
3, 1)

 -n_lien( 4, 6) f( 4) LLS( 4) w_subset( 4, 6)           +n_lien( 4, 6) f( 4) LLS( 6) w_subset(
6, 4)           avec n_lien( 4, 6) =  2
         equal to :
 -n_lien( 4, 6) f( 4) LLS( 4) w_couche( 1, 3)           +n_lien( 4, 6) f( 4) LLS( 6) w_couche(
3, 1)

 -n_lien( 4, 5) f( 4) LLS( 4) w_subset( 4, 5)           +n_lien( 4, 5) f( 4) LLS( 5) w_subset(
5, 4)           avec n_lien( 4, 5) =  1
         equal to :
 -n_lien( 4, 5) f( 4) LLS( 4) w_couche( 1, 2)           +n_lien( 4, 5) f( 4) LLS( 5) w_couche(
2, 1)

 -n_lien( 4, 2) f( 4) LLS( 4) w_subset( 4, 2)           +n_lien( 4, 2) f( 4) LLS( 2) w_subset(
2, 4)           avec n_lien( 4, 2) =  1
         equal to :
 -n_lien( 4, 2) f( 4) LLS( 4) w_couche( 1, 2)           +n_lien( 4, 2) f( 4) LLS( 2) w_couche(
2, 1)

 -n_lien( 4, 1) f( 4) LLS( 4) w_subset( 4, 1)           +n_lien( 4, 1) f( 4) LLS( 1) w_subset(
1, 4)           avec n_lien( 4, 1) =  2
         equal to :
 -n_lien( 4, 1) f( 4) LLS( 4) w_couche( 1, 1)           +n_lien( 4, 1) f( 4) LLS( 1) w_couche(
1, 1)

 -n_lien( 4, 4) f( 4) LLS( 4) w_subset( 4, 4)           +n_lien( 4, 4) f( 4) LLS( 4) w_subset(
4, 4)           avec n_lien( 4, 4) =  2
         equal to :
```

```
 -n_lien( 4, 4) f( 4) LLS( 4) w_couche( 1, 1)           +n_lien( 4, 4) f( 4) LLS( 4) w_couche(
1, 1)

 -n_lien( 4, 0) f( 4) LLS( 4) w_subset( 4, 0)           +n_lien( 4, 0) f( 4) LLS( 0) w_subset(
0, 4)           avec n_lien( 4, 0) =  1
         equal to :
 -n_lien( 4, 0) f( 4) LLS( 4) w_couche( 1, 0)           +n_lien( 4, 0) f( 4) LLS( 0) w_couche(
0, 1)
```

where f(i) and LL(i) stand for $f_i$ and $LL_i^+$ which were introduced in the main section. All the terms can be hand-summed and recast with the standard notations into Eq. (B1) of Appendix B.

The contributions to the multiplicative coefficient of $LL_4^+$ in the transport equation are printed in the output file named 'FCC_transequa_coeff_04' as:

```
 recherche de tous les termes de bilan en facteur de LL( 4)

-n_lien( 4, 0) w_subset( 4, 0) f( 4)                                        avec n_lien( 4, 0) =  1
-n_lien( 4, 1) w_subset( 4, 1) f( 4)                                        avec n_lien( 4, 1) =  2
-n_lien( 4, 2) w_subset( 4, 2) f( 4)                                        avec n_lien( 4, 2) =  1
-n_lien( 4, 4) w_subset( 4, 4) f( 4)                                        avec n_lien( 4, 4) =  2
                                        +n_lien( 4, 4) w_subset( 4, 4) f( 4)
-n_lien( 4, 5) w_subset( 4, 5) f( 4)                                        avec n_lien( 4, 5) =  1
                                        +n_lien( 4, 5) w_subset( 4, 5) f( 5)
-n_lien( 4, 6) w_subset( 4, 6) f( 4)                                        avec n_lien( 4, 6) =  2
                                        +n_lien( 4, 6) w_subset( 4, 6) f( 6)
-n_lien( 4, 7) w_subset( 4, 7) f( 4)                                        avec n_lien( 4, 7) =  2
                                        +n_lien( 4, 7) w_subset( 4, 7) f( 7)
-n_lien( 4, 8) w_subset( 4, 8) f( 4)                                        avec n_lien( 4, 8) =  1
                                        +n_lien( 4, 8) w_subset( 4, 8) f( 8)
```

which can be summed and recast into the first term in parentheses of Eq. (B6) of Appendix B.

The outward and inward contributions are merged together to make up the coefficient of $LL_4^+$ in the first equation of the system; they are printed in the file 'FCC_syslin_coeff_04_04' as:

```
coefficient a_tot( 1, 1)
 ingredients entrant dans c_tot( 4, 4) :

 + 1.d0
 + n_lien( 4, 8) fifj( 4, 4) w_subset( 4, 8)          avec n_lien( 4, 8) =  1
 + n_lien( 4, 7) fifj( 4, 4) w_subset( 4, 7)          avec n_lien( 4, 7) =  2
 + n_lien( 4, 6) fifj( 4, 4) w_subset( 4, 6)          avec n_lien( 4, 6) =  2
 + n_lien( 4, 5) fifj( 4, 4) w_subset( 4, 5)          avec n_lien( 4, 5) =  1
 + n_lien( 4, 2) fifj( 4, 4) w_subset( 4, 2)          avec n_lien( 4, 2) =  1
 + n_lien( 4, 1) fifj( 4, 4) w_subset( 4, 1)          avec n_lien( 4, 1) =  2
 + n_lien( 4, 4) fifj( 4, 4) w_subset( 4, 4)          avec n_lien( 4, 4) =  2
 + n_lien( 4, 0) fifj( 4, 4) w_subset( 4, 0)          avec n_lien( 4, 0) =  1

 ingredients entrant dans d_tot( 4, 4) :
 - n_lien( 4, 4) fifj( 4, 4) w_subset( 4, 4)          avec n_lien( 4, 4) =  2
 - n_lien( 5, 4) fifj( 4, 5) w_subset( 4, 5)          avec n_lien( 5, 4) =  4
 - n_lien( 6, 4) fifj( 4, 6) w_subset( 4, 6)          avec n_lien( 6, 4) =  1
 - n_lien( 7, 4) fifj( 4, 7) w_subset( 4, 7)          avec n_lien( 7, 4) =  2
 - n_lien( 8, 4) fifj( 4, 8) w_subset( 4, 8)          avec n_lien( 8, 4) =  1
```

where fifj(i,j) stands for the integral $\frac{1}{n_i\, V_{ZB}} \int \frac{f_i(k) f_j(k)}{D_0} d_3k$. All the terms can be summed and recast into the top-left coefficient displayed in Table B2 of Appendix B.

The integrals are easily calculated once for all for a given lattice symmetry; but their numbering depends on the subsets which are introduced for the calculation. In the above quoted example of BCC lattice, the function associated to the subset $(2,2,2)_5$ is called $F_6$ , $F_8$ , $F_{10}$ , $F_{12}$ if the interaction range extends up to 1st, 2nd, 7th, 10th neighbour shell respectively. This is the reason why an alternative notation was introduced, which does not depend on the interaction range as explained in the main section.

The numerical values of integrals are obtained through a superposition of gaussian quadratures on the 3 axes, as already explained in Appendix B.

The subroutine containing the Fortran instructions defining the functions $F_i$ is generated automatically to avoid the building of a sparse matrix.

All the operations are monitored through an executable using shell commands (the Fortran code is available at the author on request).

## Appendix D: Relationships between lattice integrals

As explained in the main section, the correlation factors obtained with model I do not depend on the interaction range.
The source of this surprising fact must be traced back in the fortuitous cancelation of off-diagonal coefficients. The latter is due to the existence of linear relationships between lattice integrals $fifj(k,l) \equiv fifj_{k_1k_2k_3\times l_1l_2l_3}$ appearing in the coefficients.

For the FCC lattice with an interaction ranging up to 5th neighbour shell, we recall hereafter the numbers attached to the relevant subsets. Each subset number is associated with the triplet of coordinates of its representative site, together with a lower subscript indicating the number of the neighbour shell to which it belongs for sake of convenience:
8≡$(101)_1$ ; 9≡$(200)_2$ ; 10≡$(112)_3$ ; 11≡$(211)_3$ ; 12≡$(202)_4$ ; 13≡$(103)_5$ ; 14≡$(301)_5$ ; 15≡$(222)_6$ ; 16≡$(123)_7$ ;17≡$(213)_7$ ; 18≡$(312)_7$ ; 19≡$(400)_8$ ; 20≡$(114)_9$ ; 21≡$(303)_9$ ; 22≡$(411)_9$ ; 23≡$(204)_{10}$ ; 24≡$(402)_{10}$ .

In the ouput files named ‘FCC_syslin_coeff_xx_yy ‘, the only pairs (xx,yy) to be considered are (08,09) ; (08,10) ; (08,11) ; (08,12). Indeed starting from the 1rst neighbour shell, a single jump gives only access to 2nd, 3rd and 4th neighbour shells but not beyond.

The brute content of ‘FCC_syslin_coeff_08_09’ is displayed below in its standard form:

```
coefficient renumerote a_tot( 1, 2)
 ingredients entrant dans c_tot( 8, 9) :

 + n_lien( 9,14) fifj( 8, 9) w_subset( 9,14)  avec n_lien( 9,14) =  4  fifj( 8, 9) =
1.62484092692833D-02  et w_subset( 9,14) =  1.5304D+00
 + n_lien( 9,11) fifj( 8, 9) w_subset( 9,11)  avec n_lien( 9,11) =  4  fifj( 8, 9) =
1.62484092692833D-02  et w_subset( 9,11) =  1.5304D+00
 + n_lien( 9, 8) fifj( 8, 9) w_subset( 9, 8)  avec n_lien( 9, 8) =  4  fifj( 8, 9) =
1.62484092692833D-02  et w_subset( 9, 8) =  1.5304D+00

 ingredients entrant dans d_tot( 8, 9) :
 - n_lien( 8, 9) fifj( 8, 8) w_subset( 9, 8)  avec n_lien( 8, 9) =  1  fifj( 8, 8) =
1.22680066019576D-01  et w_subset( 9, 8) =  1.5304D+00
 - n_lien(11, 9) fifj( 8,11) w_subset( 9,11)  avec n_lien(11, 9) =  1  fifj( 8,11) =
4.37490822468529D-02  et w_subset( 9,11) =  1.5304D+00
 - n_lien(14, 9) fifj( 8,14) w_subset( 9,14)  avec n_lien(14, 9) =  1  fifj( 8,14) =
2.85517629649070D-02  et w_subset( 9,14) =  1.5304D+00

 valeur finale a_tot( 1, 2) =     9.73110481083950D-14
```

It shows that the coefficient a_tot(1,2) is vanishingly small because:

$$12\,fifj(8,9) = fifj(8,8) + fifj(8,11) + fifj(8,14)$$
$$\Rightarrow 12\,fifj_{101\times200} = fifj_{101\times101} + fifj_{101\times211} + fifj_{101\times301}$$

with the definitions :

$$F_{101} = 4s_x(c_y + c_z) \qquad F_{200} = 2s_{2x} = 4s_xc_x \qquad F_{211} = 8s_xc_xc_yc_z \qquad F_{301} = 4s_{3x}(c_y + c_z)$$

This identity is easy to establish. The numerator of the integral on the left side is :

$$12F_{101}F_{200} = 12\times4s_x(c_y + c_z)\times4s_xc_x = 192s_x^2c_x(c_y + c_z)$$

The numerator of the integral on the right side is given by :

$$F_{101}(F_{101} + F_{211} + F_{301}) = 4s_x(c_y + c_z)\times16s_xc_x(c_yc_z + c_xc_y + c_xc_z) = 16s_x^2c_x(c_y + c_z)(12 - D_0)$$
$$= 192s_x^2c_x(c_y + c_z) - 16s_x^2c_x(c_y + c_z)D_0$$

The difference between the two numerators is the second term of the preceding line: it is proportional to the denominator $D_0$. The remaining part to be integrated is thus a polynomial with odd power of trigonometric functions. Hence

$$\frac{1}{VZB}\int_{VZB}\frac{F_{101}\{12F_{200}-(F_{101}+F_{211}+F_{301})\}}{D_0}d_3k=\frac{1}{VZB}\int_{VZB}16s_x^2c_x(c_y+c_z)d_3k=0$$

Further examination of the ingredients entering the vanishing coefficients a_tot(1,3), a_tot(1,4), a_tot(1,5) in the output files 'FCC_syslin_coeff_08_10', 'FCC_syslin_coeff_08_11', 'FCC_syslin_coeff_08_12' respectively shows that the above situation is still encountered, which yields three other identities. For all of them it can be checked that the difference between the two numerators can be divided by $D_0$ and that the remaining term contains always odd powers of $c_y$ and $c_z$:

$$12fifj(8,10)=2fifj(8,8)+fifj(8,10)+fifj(8,11)+fifj(8,12)+fifj(8,13)+fifj(8,15)$$
$$\Rightarrow 12fifj_{101\times112}=2fifj_{101\times101}+fifj_{101\times112}+fifj_{101\times211}+fifj_{101\times202}+fifj_{101\times103}+fifj_{101\times222}$$
$$12fifj(8,11)=2fifj(8,8)+4fifj(8,9)+fifj(8,10)+2fifj(8,12)+2fifj(8,14)+fifj(8,15)+fifj(8,18)$$
$$\Rightarrow 12fifj_{101\times211}=2fifj_{101\times101}+4fifj_{101\times200}+fifj_{101\times112}+2fifj_{101\times202}+2fifj_{101\times301}+fifjj_{101\times222}+fifj_{101\times312}$$
$$12fifj(8,12)=fifj(8,8)+fifj(8,10)+2fifj(8,11)+fifj(8,13)+fifj(8,14)+fifj(8,17)+fifj(8,18)+fifj(8,21)$$
$$\Rightarrow 12fifj_{101\times202}=fifj_{101\times101}+fifj_{101\times112}+2fifj_{101\times211}+fifj_{101\times103}+fifj_{101\times301}+fifj_{101\times213}+fifj_{101\times312}+fifj_{101\times303}$$

For the BCC lattice the relationships to be established amount to five since the 5[th] neighbour shell can be reached in one jump from the 1[st] neighbour shell. The numbering of subsets is recalled hereafter:
5≡$(111)_1$ ; 6≡$(200)_2$ ; 7≡$(202)_3$ ; 8≡$(113)_4$ ; 9≡$(311)_4$ ; 10≡$(222)_5$ ; 11≡$(400)_6$ ; 12≡$(133)_7$ ; 13≡$(313)_7$ ; 14≡$(204)_8$ ; 15≡$(402)_8$ ; 16≡$(224)_9$ ; 17≡$(422)_9$ ; 18≡$(115)_{10}$ ; 19≡$(333)_{10}$ ; 20≡$(511)_{10}$ .

The output files to be examined are 'BCC_syslin_coeff_xx_yy' with (xx, yy)= (5,6), (5,7), (5,8), (5,9), (5,10), which yields the five identities of interest:

$$8fifj(5,6)=fifj(5,5)+fifj(5,9)$$
$$\Rightarrow 8fifj_{111\times200}=fifj_{111\times111}+fifj_{111\times311}$$
$$8fifj(5,7)=2fifj(5,5)+fifj(5,8)+2fifj(5,9)+fifj(5,13)$$
$$\Rightarrow 8fifj_{111\times202}=2fifj_{111\times111}+fifj_{111\times113}+2fifj_{111\times311}+fifj_{111\times313}$$
$$8fifj(5,8)=2fifj(5,7)+2fifj(5,10)+2fifj(5,14)+fifj(5,16)$$
$$\Rightarrow 8fifj_{111\times113}=2fifj_{111\times202}+fifj_{111\times222}+2fifj_{111\times204}+fifj_{111\times224}$$
$$8fifj(5,9)=4fifj(5,6)+2fifj(5,7)+fifj(5,10)+4fifj(5,11)+2fifj(5,15)+fifj(5,17)$$
$$\Rightarrow 8fifj_{111\times311}=4fifj_{111\times200}+2fifj_{111\times202}+fifj_{111\times222}+4fifj_{111\times400}+2fifj_{111\times402}+fifj_{111\times422}$$
$$8fifj(5,10)=fifj(5,5)+fifj(5,8)+fifj(5,9)+fifj(5,12)+fifj(5,13)+fifj(5,19)$$
$$\Rightarrow 8fifj_{111\times222}=fifj_{111\times111}+fifj_{111\times113}+fifj_{111\times311}+fifj_{111\times133}+fifj_{111\times313}+fifj_{111\times333}$$

Of course, similar linear relationships do exist between integrals of higher indices; but they can no longer alter the value of the return probability on the first neighbour shell.